\documentclass[a4paper,fleqn,usenatbib]{mnras}

\usepackage[T1]{fontenc}
\usepackage{ae,aecompl}
\usepackage{graphicx} 
\usepackage{amsmath} 
\usepackage{amssymb}     
\usepackage{subfigure}
\usepackage{multirow}
\usepackage{array}
\usepackage{booktabs}
\usepackage{multirow}
\usepackage{makecell}

\usepackage{hyperref}
\hypersetup{hypertex=true, colorlinks=true, linkcolor=blue, anchorcolor=blue, citecolor=blue}

\title[Gas-phase hydrogenation of PAH cations]
{Gas-phase hydrogenation of large, astronomically relevant PAH cations}
\author[Hua et al.]{
Lijun Hua,$^{1}$
Xiaoyi Hu,$^{2}$
Junfeng Zhen,$^{3}$\thanks{E-mail: jfzhen@ustc.edu.cn}
Xuejuan Yang,$^{1}$\thanks{E-mail: xjyang@xtu.edu.cn}
\\
$^{1}$ Hunan Key Laboratory for Stellar and Interstellar Physics and School of Physics and Optoelectronics, \\Xiangtan University, Hunan 411105, China \\
$^{2}$ Department of Chemical Physics, University of Science and Technology of China, 96 Jinzhai RD., Hefei, Anhui 230026, China\\
$^{3}$ Institute of Deep Space Sciences, Deep Space Exploration Laboratory, Hefei 230026, China\\
}

\begin{document}
\label{firstpage}
\pagerange{\pageref{firstpage}
\pageref{lastpage}}
\maketitle

\begin{abstract}

To investigate the gas-phase hydrogenation processes of large, astronomically relevant cationic polycyclic aromatic hydrocarbon (PAH) molecules under the interstellar environments, the ion-molecule collision reaction between six PAH cations and H-atoms is studied. The experimental results show that the hydrogenated PAH cations are efficiently formed, and no even-odd hydrogenated mass patterns are observed in the hydrogenation processes. The structure of newly formed hydrogenated PAH cations and the bonding energy for the hydrogenation reaction pathways are investigated with quantum theoretical calculations. The exothermic energy for each reaction pathway is relatively high, and the competition between hydrogenation and dehydrogenation is confirmed. From the theoretical calculation, the bonding ability plays an important role in the gas-phase hydrogenation processes. The factors that affect the hydrogenation chemical reactivity are discussed, including the effect of carbon skeleton structure, the side-edged structure, the molecular size, the five- and six-membered C-ring structure, the bay region structure, and the neighboring hydrogenation. The IR spectra of hydrogenated PAH cations are also calculated. These results we obtain once again validate the complexity of hydrogenated PAH molecules, and provide the direction for the simulations and observations under the coevolution interstellar chemistry network. We infer that if we do not consider other chemical evolution processes (e.g., photo-evolution), then the hydrogenation states and forms of PAH compounds are intricate and complex in the interstellar medium (ISM).

\end{abstract}



\section{Introduction}
\label{sec:intro}

\begin{table*}
      \centering
      \caption{The molecular geometry of PAH molecules. The different edge structures are highlighted with different colors: solo (purple), duo (red), trio (green), and quarto (blue). Edge bonds without hydrogens attached correspond to bay regions (black), which can take different forms.}
      \label{tab:chap:table1}
      \renewcommand\arraystretch{2}%
      \setlength{\tabcolsep}{1.2mm}
      \begin{tabular}[c]{ccccccc}
             \toprule
             {PAHs} & {structures} & {solo} & {duo} & {trio} & {quarto} & {bay regons}\\
             \midrule[0.8pt]
             Ovalene ($\rm{C_{32}H_{14}}$, m/z=398) & \begin{minipage}[b]{0.19\columnwidth}
                    \centering
                    \raisebox{-.5\height}{\includegraphics[width=\linewidth]{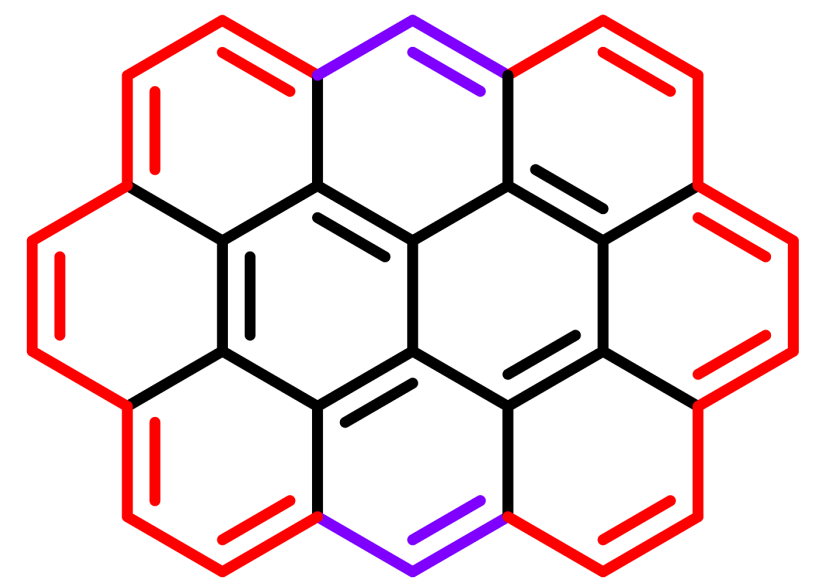}}
             \end{minipage} \ & 2 & 6 & 0 & 0 & 0\\
             \midrule[0.3pt]%
             Periflanthene ($\rm{C_{32}H_{16}}$, m/z=400) & \begin{minipage}[b]{0.27\columnwidth}
                    \centering
                    \raisebox{-.5\height}{\includegraphics[width=\linewidth]{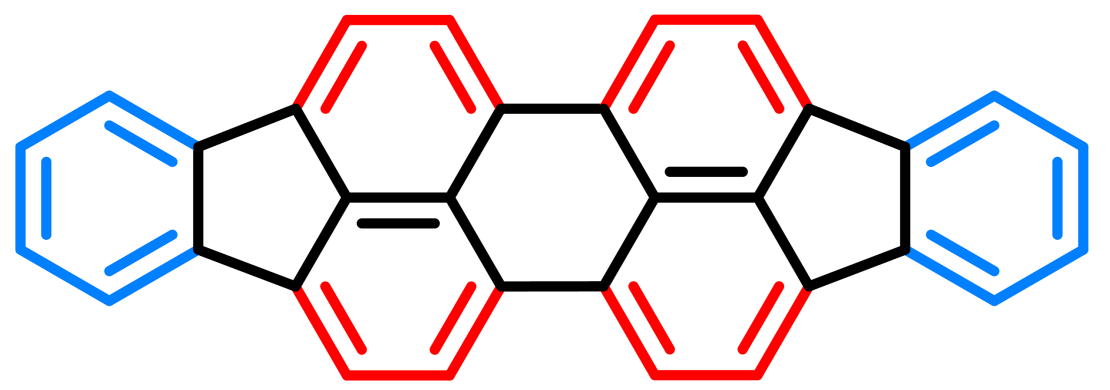}}
             \end{minipage} \ & 0 & 4 & 0 & 2 & 6 \\
             \midrule[0.3pt]%
             Tri-benzo-peropyrene (TBP, $\rm{C_{34}H_{16}}$, m/z=424) & \begin{minipage}[b]{0.215\columnwidth}
                    \centering
                    \raisebox{-.5\height}{\includegraphics[width=\linewidth]{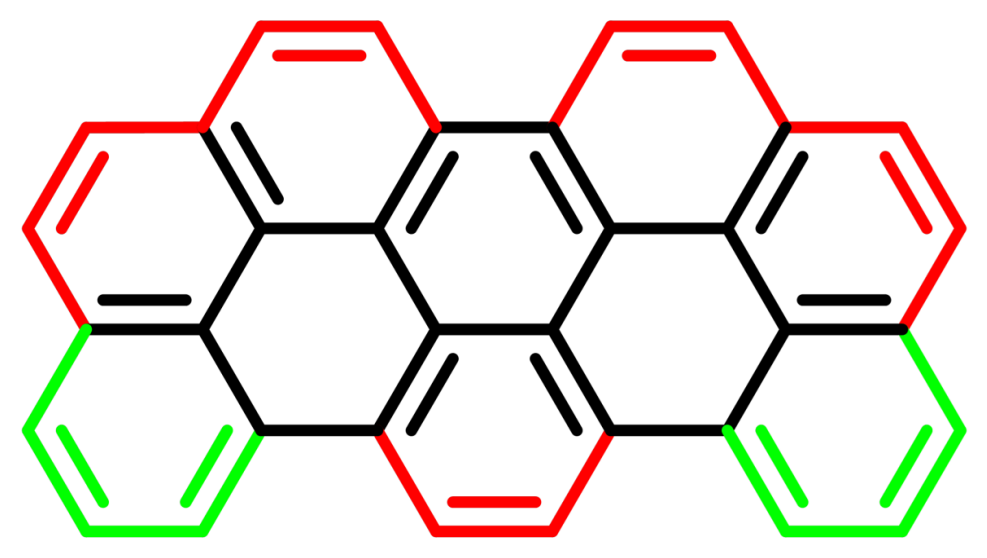}}
             \end{minipage} \ & 0 & 5 & 2 & 0 & 3 \\
             \midrule[0.3pt]%
             Tri-benzo-naphtho-pero-pyrene (TNPP, $\rm{C_{40}H_{18}}$, m/z=498) & \begin{minipage}[b]{0.21\columnwidth}
                    \centering
                    \raisebox{-.5\height}{\includegraphics[width=\linewidth]{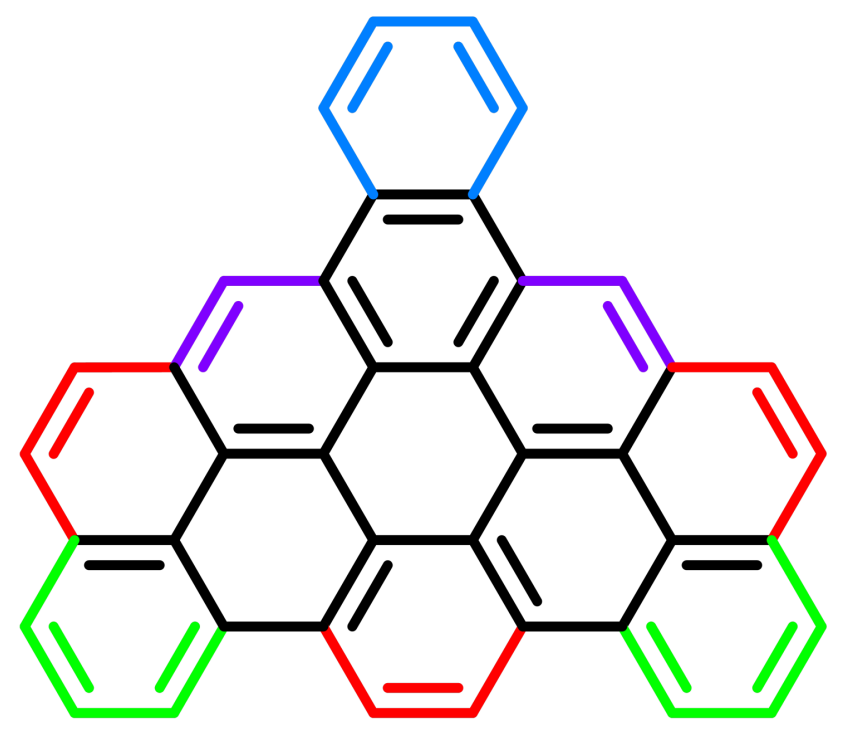}}
             \end{minipage} \ & 2 & 3 & 2 & 1 & 4 \\
             \midrule[0.3pt]%
             Hexa-benzo-coronene (HBC, $\rm{C_{42}H_{18}}$, m/z=522) & \begin{minipage}[b]{0.21\columnwidth}
                    \centering
                    \raisebox{-.5\height}{\includegraphics[width=\linewidth]{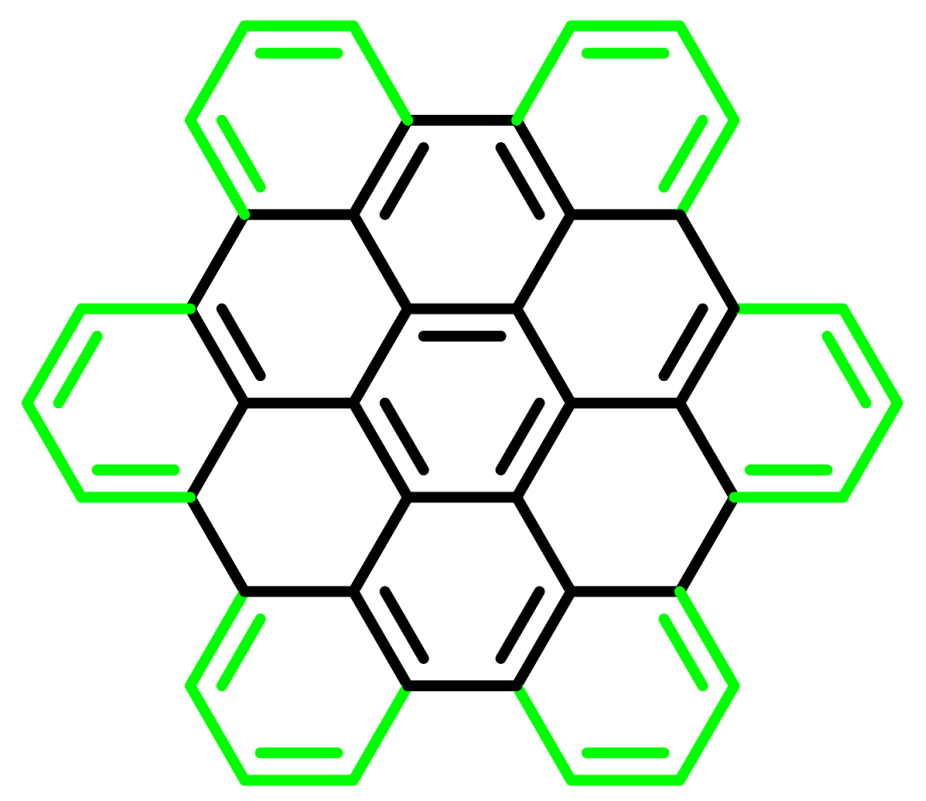}}
             \end{minipage} \ & 0 & 0 & 6 & 0 & 6 \\
             \midrule[0.3pt]%
             Dicoronylene (DC, $\rm{C_{48}H_{20}}$, m/z=598) & \begin{minipage}[b]{0.29\columnwidth}
                    \centering
                    \raisebox{-.5\height}{\includegraphics[width=\linewidth]{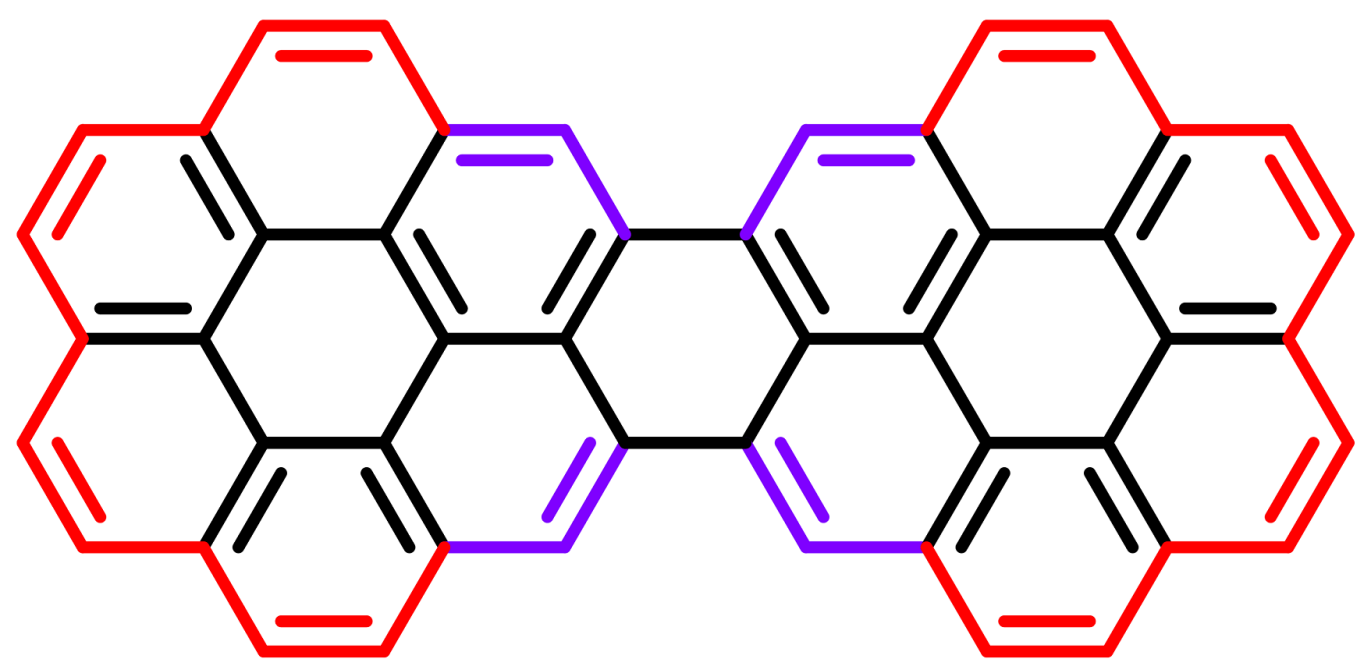}}
             \end{minipage} \ & 4 & 8 & 0 & 0 & 2 \\
             \bottomrule
      \end{tabular}
\end{table*}

PAH molecules and their derivatives are abundant and ubiquitous in many astronomical sources and are expected to account for $\sim$ 10\% of cosmic carbon in the ISM \citep{sel84, all89, pug89}. The assignment of the aromatic infrared bands (AIBs) to PAHs has widely been discussed and accepted \citep{leg84,all85,all89,pee04}. The prominent 3.3 and 3.4 $\mu$m feature is associated with aromatic and aliphatic CH stretches \citep{sch93,ber96,pen02}, and the 6.2 and 7.7 $\mu$m bands are linked to CC vibrations of aromatic rings. The 8.6 $\mu$m and 11.2 $\mu$m are assigned to CH in- and out-of-plane bending and \citep{pee04,tie08}.And longer wavelength bands (>16 $\mu$m) are assigned to CCC bending \citep{all89}. The relative strength of these bands depends on the chemical properties (e.g., size, structure, and charge states) of PAH molecules \citep{all99,dra01} and their physical environment (e.g., the intensity and hardness of starlight illuminating PAHs, electron density, and gas temperature) \citep{bak94,wei01,yang2021}.

As a significant class of large organic interstellar molecules, PAHs play an important role in the energy and ionization balance of the ISM \citep[and references therein]{tie08, berfo22}, and catalyze the formation of molecular H$_2$ in photo-dissociation regions (PDRs; \cite{bos12}). Interstellar chemical-evolution processes, including collision reactions (e.g., hydrogenation), photo-fragmentation, and ionization, further modify the states and forms of these PAH compounds in interstellar environments that various types of functional groups may be acquired and functionalize, such as H/D, methyl (-CH$_3$), vinyl (-CHCH$_2$), methoxy (-OCH$_3$), amino (-NH$_2$), cyano/isocyano (-CN, -NC), acid (-COOH), and hydroxyl (-OH) \citep[and references therein]{hol99,tie13}.

PAHs and their associated derivate clusters may be formed through ion-molecule reactions in interstellar space where PAH and other coexist atoms or molecules \citep{pet93,boh16,zhen2018,zhen2019}. Hydrogen is the most abundant element in the universe, and there are some regions in the interstellar space where hydrogen (H) atoms are abundant, such as PDRs. PDRs provide a natural laboratory for the study of the interaction of interstellar UV radiation, hydrogen atoms, hydrogen ions, and other molecules coexisting with carbonaceous species \citep{pet05,rapacioli05,hol99,mic10}. Observations with the Spitzer Space Telescope and the Herschel Space Observatory of the prototypical PDR, NGC 7023, have revealed the important role of hydrogenation processes of interstellar PAHs \citep{and16}.

When PAHs are in this high hydrogen density environment and are continuously bombarded by H-atoms, there may be excess H-atoms on the periphery \citep[and references therein]{and16}, which are also known as super-hydrogenated PAHs. The interaction between PAHs and H-atoms has been studied theoretically \citep{cas94,bau98, E. Rauls2008, J. A. Sebree2010,yang20a} and experimentally \citep{Snow et al.1998,thr12,caz16,caz19,jen19}, and these studies indicate that super-hydrogenation of PAH cations is possible. In addition, the IR vibrational spectra of mono-deuterated PAHs have also been performed by theoretical calculations, and the intrinsic band strength of the 4.4 $\mu$m C-D stretch determined \citep{yang20b,yang2021}. Furthermore, in a combined experimental and theoretical study, coronene (C$_{24}$H$_{12}$) hydrogenation follows a well-defined sequence of hydrogenation sites that has later been confirmed by means of infrared spectroscopy \citep{caz16,caz19}. Moreover, the gas-phase hydrogen/deuterium exchange on large, astronomically relevant cationic PAHs (hexa-benzo-coronene cation, $\rm{[C_{42}H_{18}]}$$^+$) is also studied \citep{zhang22}.

To understand the reaction between PAH cations and H-atoms and further investigate the coevolution network of the interstellar PAH chemistry \citep{tie13,zhang23}, in this work, by tracking the hydrogenation processes of PAH compounds, we present an exploration study of the chemical reactivity of six large, astronomically relevant PAH cations with H-atoms. Furthermore, to investigate the formation mechanism of hydrogenated PAH cations, the experimental results are illustrated together with the results of the theoretical chemistry calculation.

In this work, ovalene ($\rm{C_{32}H_{14}}$), periflanthene ($\rm{C_{32}H_{16}}$), tri-benzo-peropyrene (TBP, $\rm{C_{34}H_{16}}$), tribenzo-naphtho-pero-pyrene (TNPP, $\rm{C_{40}H_{18}}$), hexa-benzo-coronene (HBC, $\rm{C_{42}H_{18}}$), dicoronylene (DC, $\rm{C_{48}H_{20}}$) are selected, and the molecular geometry and edge structure information are summarized in Table 1. These selected PAHs can be used as a prototypical example for large(r) PAHs \citep{and2015,cro16}, and its size, from 32 to 48 C-number, that lies within the astrophysically relevant range \citep{kok08,ste11,gre11}, and also allows us to unambiguously identify the hydrogenation processes for PAHs with all types of molecular geometry.

\section{Experimental and calculation results}
\label{sec:results}

The experiment was performed on the apparatus equipped with a quadrupole ion trap and reflection TOF mass spectrometer, together with a heat-able oven for the natural gas-phase PAH species (provided by Kentax, with a purity greater than 99.5 \%) \citep{zhang23} (details in Appendix A). Evaporated natural PAH molecules were ionized using electron impact ionization ($\sim$ 82 eV) and transported into the ion trap via an ion-gate. A hydrogen atom source is installed to produce the H-atoms (HABS; MBE-Komponenten GmbH; \cite{tsc98}). Based on the operating conditions, the H-atoms were expected to have a flux of $\sim$ 4.0 * 10$^{14}$ H-atoms cm$^{-2}$s$^{-1}$, and the volume density of H-atoms is $\sim$ 5*10$^9$ cm$^{-3}$. The resulting mass spectrum of large, astronomically relevant hydrogenated cationic PAHs is obtained and presented in the left panel of Figures 1-6.

To illustrate the hydrogenation mechanism of large, astronomically relevant PAH molecules, we theoretically study the reaction pathways between PAH cations and H-atoms. The theoretical calculations were performed based on density functional theory (DFT) with the hybrid density functional B3LYP \citep{bec92,lee88}, as implemented in the Gaussian 16 program \citep{fri16}. The basis set of the 6-311++G** was selected and used for all these systems. To account for the intermolecular interactions, the dispersion-correction (D3, \citep{gri11}) is considered for each system. All species' geometries were optimized at the local minimum of their potential energy surface in the calculation. Furthermore, the zero-point vibrational energy (ZPVE) was obtained from the frequency calculation to correct the molecular energy.

In addition, we note that due to the higher reactivity of H/D atoms with PAH cations, the energy barrier in the accretion of H/D atoms on PAH cations is very small, in the range of $\sim$10 to 30 meV, which have two orders of magnitude smaller than the exothermic energy ($\sim$1.0 to 3.0 eV) \citep{caz16}. In other words, the attachment energy barriers are almost vanished, which we suppose it may be not play an important role in the hydrogenation process. In addition, we suppose the collisional energy between the PAH cations and H-atoms is not very high ($\sim$ 0.1 to 0.3 eV). So only the exothermic energy for the reaction pathways needs to be considered in the theoretical calculation \citep{caz16,sch20,zhang23}.
	
Furthermore, for the ion-atom collision reactions that under the gas phase conditions, we consider the reaction collision can be happened if the exothermic energy is around or higher than $\sim$ 1.0 eV. Given this, from our obtained theoretical calculation results, the exothermic energy at each step is all around or higher than $\sim$ 1.0 eV, so we regard the ion-atom collision reaction rate between PAH cations and H atoms as the rate of reaction in the gas phase \citep{sch20,zhang23}.

The obtained calculation results of hydrogenated cationic large, astronomically relevant PAHs are presented in the right panel of Figures 1-6 and Table 2-3. For all these PAH cations and hydrogenated PAH cations, the doublet spin multiplicity is considered in the calculations.

\begin{figure*}
      \subfigure{
             \begin{minipage}[t]{0.5\linewidth}
                    \centering
                    \includegraphics[width=\linewidth]{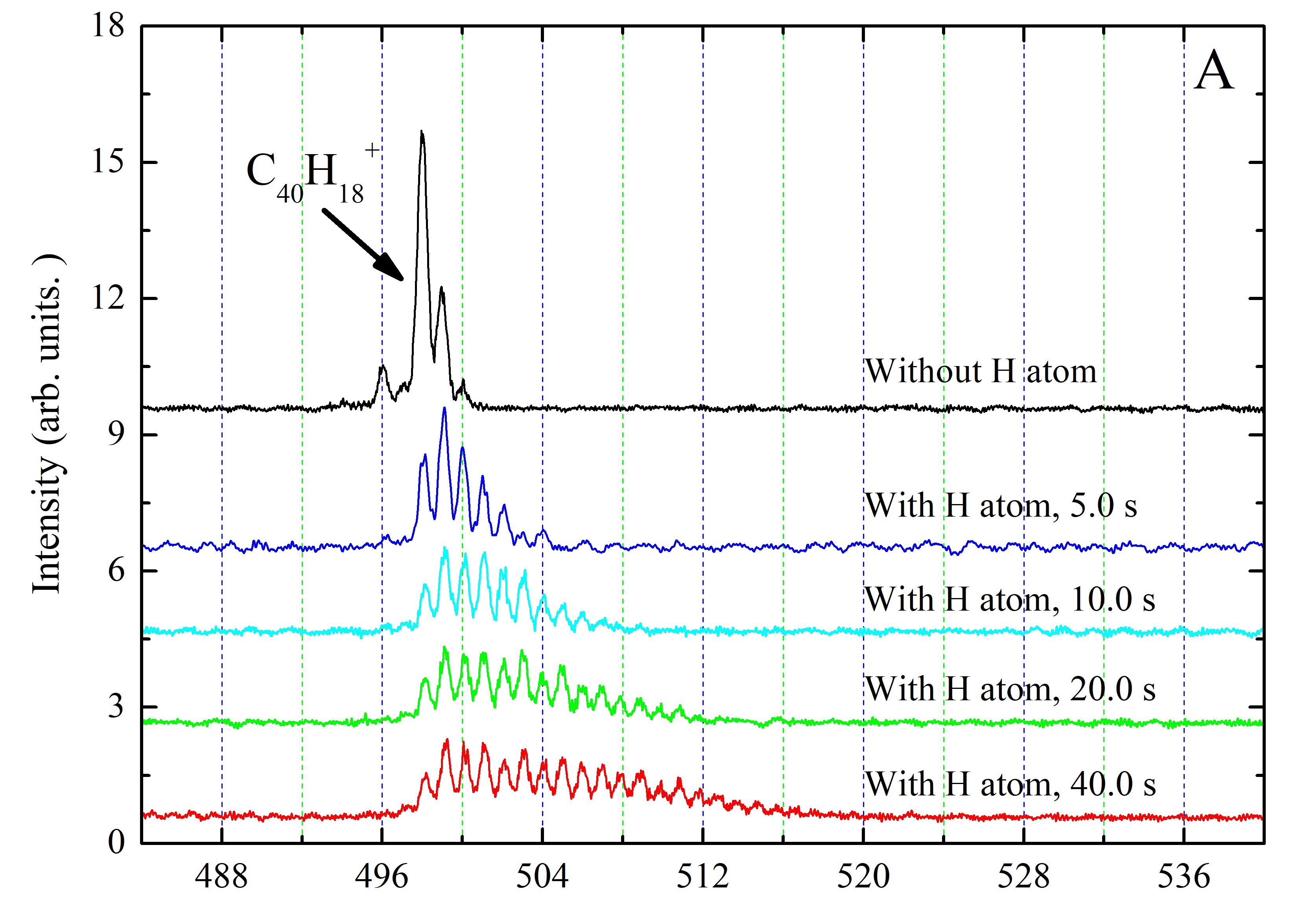}\\
                    \vspace{0.1cm}
                    \includegraphics[width=\linewidth]{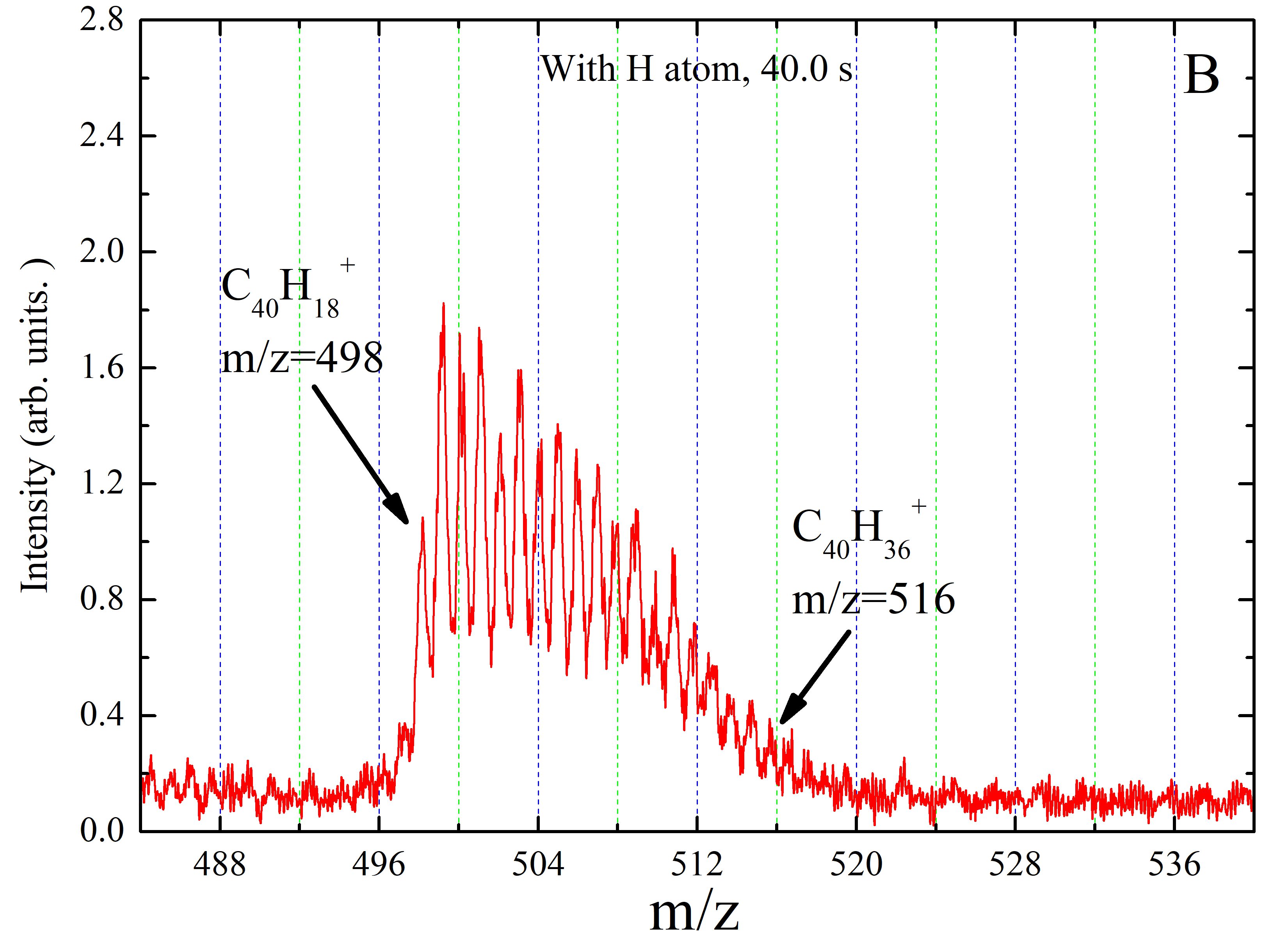}
                    \vspace{0.02cm}
             \end{minipage}%
      }%
      \subfigure{
             \begin{minipage}{0.4\linewidth}
                    \centering
                    \renewcommand\arraystretch{1.2}
                    \setlength{\tabcolsep}{8mm}
                    \begin{tabular}{|c|c|}
                           \hline
                           \multicolumn{2}{|c|}{$\rm{[C_{40}H_{18}]}^+$} \\ \hline
                           \specialrule{0em}{0pt}{0.25pt}
                           \multicolumn{2}{|c|}{
                                  \begin{minipage}[b]{0.25\columnwidth}
                                         \centering
                                        \raisebox{-.5\height}{\includegraphics[width=\linewidth]{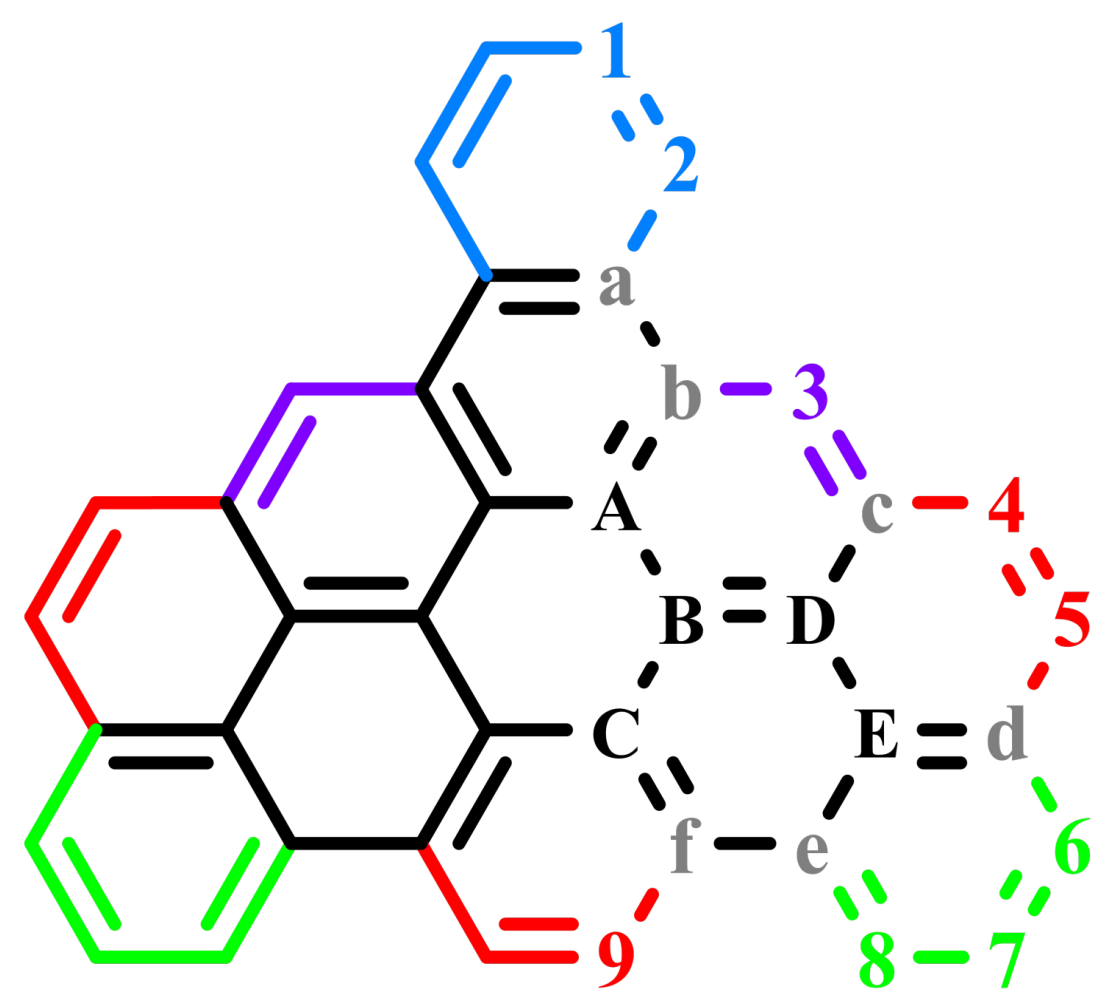}}
                           \end{minipage}} \\ \hline
                           \multicolumn{2}{|c|}{\textbf{Outer carbon sites}} \\ \hline
                           {$\rm{[C_{40}H_{18}+H(1)]}^+$} & -1.53 eV \\ \hline
                           {$\rm{[C_{40}H_{18}+H(2)]}^+$} & -1.54 eV \\ \hline
                           {$\rm{[C_{40}H_{18}+H(3)]}^+$} & -2.14 eV \\ \hline
                           {$\rm{[C_{40}H_{18}+H(4)]}^+$} & -1.54 eV \\ \hline
                           {$\rm{[C_{40}H_{18}+H(5)]}^+$} & -1.79 eV \\ \hline
                           {$\rm{[C_{40}H_{18}+H(6)]}^+$} & -2.10 eV \\ \hline
                           {$\rm{[C_{40}H_{18}+H(7)]}^+$} & -1.45 eV \\ \hline
                           {$\rm{[C_{40}H_{18}+H(8)]}^+$} & -2.11 eV \\ \hline
                           {$\rm{[C_{40}H_{18}+H(9)]}^+$} & -1.77 eV \\ \hline
                           \multicolumn{2}{|c|}{Inner carbon sites} \\ \hline
                           {$\rm{[C_{40}H_{18}+H(a)]}^+$} & -1.09 eV \\ \hline
                           {$\rm{[C_{40}H_{18}+H(b)]}^+$} & -1.30 eV \\ \hline
                           {$\rm{[C_{40}H_{18}+H(c)]}^+$} & -1.16 eV \\ \hline
                           {$\rm{[C_{40}H_{18}+H(d)]}^+$} & -0.93 eV \\ \hline
                           {$\rm{[C_{40}H_{18}+H(e)]}^+$} & -1.03 eV \\ \hline
                           {$\rm{[C_{40}H_{18}+H(f)]}^+$} & -1.37 eV \\ \hline
                           \multicolumn{2}{|c|}{Center carbon sites} \\ \hline
                           {$\rm{[C_{40}H_{18}+H(A)]}^+$} & -1.31 eV \\ \hline
                           {$\rm{[C_{40}H_{18}+H(B)]}^+$} & -1.19 eV \\ \hline
                           {$\rm{[C_{40}H_{18}+H(C)]}^+$} & -1.03 eV \\ \hline
                           {$\rm{[C_{40}H_{18}+H(D)]}^+$} & -1.22 eV \\ \hline
                           {$\rm{[C_{40}H_{18}+H(E)]}^+$} & -0.98 eV \\
                           \hline
                    \end{tabular}
                    \vspace{0.06cm}
             \end{minipage}%
      }%
      \centering
      \caption{Panel (A): evolution of the mass spectrum of the hydrogenated TNPP cations with increasing H-atoms exposure time of 5.0, 10.0, 20.0, and 40.0 s; panel (B): zoomed-in mass spectrum (40.0 s), revealing the presence of a highly broadened mass distribution, and the large hydrogenated TNPP cations $\rm[{C_{40}H_{18+18}}]^+$; right panel: the theoretical calculation results for [C$_{40}$H$_{18}$]$^+$ $+$ H.}
      \vspace{0.02cm}
      \label{fig1}
\end{figure*}

\begin{figure*}
      \subfigure{
             \begin{minipage}[t]{0.5\linewidth}
                    \centering
                    \includegraphics[width=\linewidth]{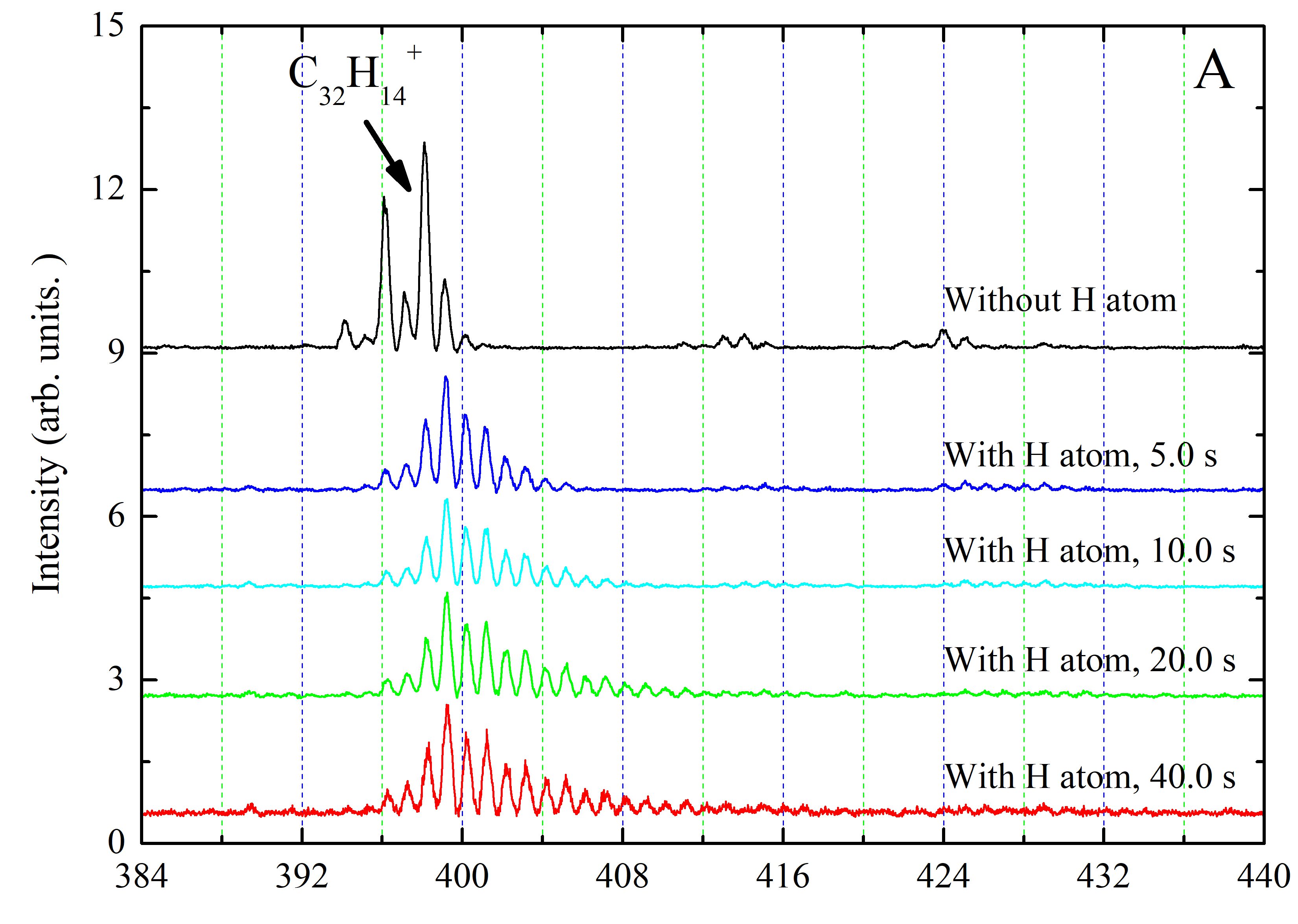}\\
                    \vspace{0.02cm}
                    \includegraphics[width=\linewidth]{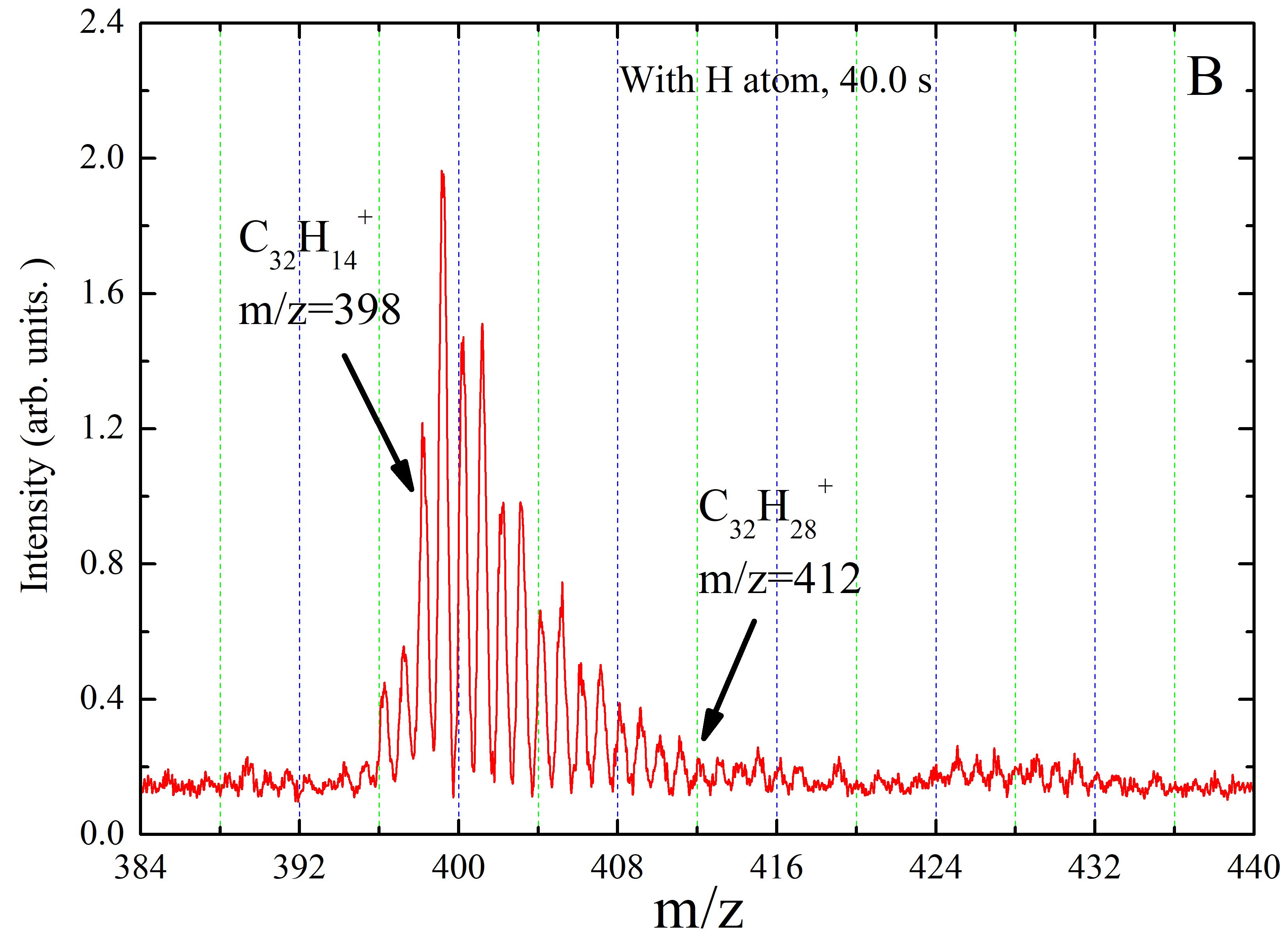}
                    \vspace{0.02cm}
             \end{minipage}%
      }%
      \subfigure{
             \begin{minipage}{0.4\linewidth}
                    \centering
                    \renewcommand\arraystretch{2.14}
                    \setlength{\tabcolsep}{8mm}
                    \begin{tabular}{|c|c|}
                           \hline
                           \multicolumn{2}{|c|}{\textbf{$\rm{[C_{32}H_{14}]}^+$}} \\ \hline
                           \specialrule{0em}{0pt}{0.25pt}
                           \multicolumn{2}{|c|}{
                                  \begin{minipage}[b]{0.4\columnwidth}
                                         \centering
                                        \raisebox{-.5\height}{\includegraphics[width=\linewidth]{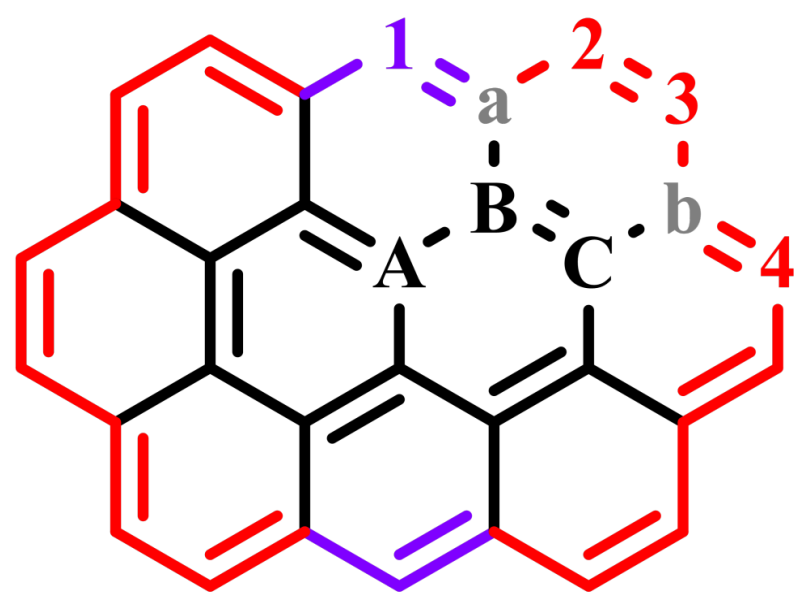}}
                           \end{minipage}} \\ \hline
                           \multicolumn{2}{|c|}{\textbf{Outer carbon sites}} \\ \hline
                           {$\rm{[C_{32}H_{14}+H(1)]}^+$} & -2.31 eV \\ \hline
                           {$\rm{[C_{32}H_{14}+H(2)]}^+$} & -1.84 eV \\ \hline
                           {$\rm{[C_{32}H_{14}+H(3)]}^+$} & -1.59 eV \\ \hline
                           {$\rm{[C_{32}H_{14}+H(4)]}^+$} & -1.82 eV \\ \hline
                           \multicolumn{2}{|c|}{\textbf{Inner carbon sites}} \\ \hline
                          {$\rm{[C_{32}H_{14}+H(a)]}^+$} & -0.94 eV \\ \hline
                           {$\rm{[C_{32}H_{14}+H(b)]}^+$} & -1.23 eV \\ \hline
                           \multicolumn{2}{|c|}{\textbf{Center carbon sites}} \\ \hline
                           {$\rm{[C_{32}H_{14}+H(A)]}^+$} & -0.98 eV \\ \hline
                           {$\rm{[C_{32}H_{14}+H(B)]}^+$} & -1.14 eV \\ \hline
                           {$\rm{[C_{32}H_{14}+H(C)]}^+$} & -0.96 eV \\
                           \hline
                    \end{tabular}
                    \vspace{0.06cm}
             \end{minipage}%
      }%
      \centering
      \caption{Panel (A): evolution of the mass spectrum of the hydrogenated ovalene cations with increasing H atom exposure time of 5.0, 10.0, 20.0, and 40.0 s; panel (B): zoomed-in mass spectrum (40.0 s), revealing the presence of a highly broadened mass distribution, and the large hydrogenated ovalene cations $\rm[{C_{32}H_{14+14}}]^+$; right panel: the theoretical calculation results for [C$_{32}$H$_{14}$]$^+$ $+$ H.
      }
      \vspace{0.02cm}
      \label{Fig2}
\end{figure*}

\begin{figure*}
      \subfigure{
             \begin{minipage}[t]{0.5\linewidth}
                    \centering
                    \includegraphics[width=\linewidth]{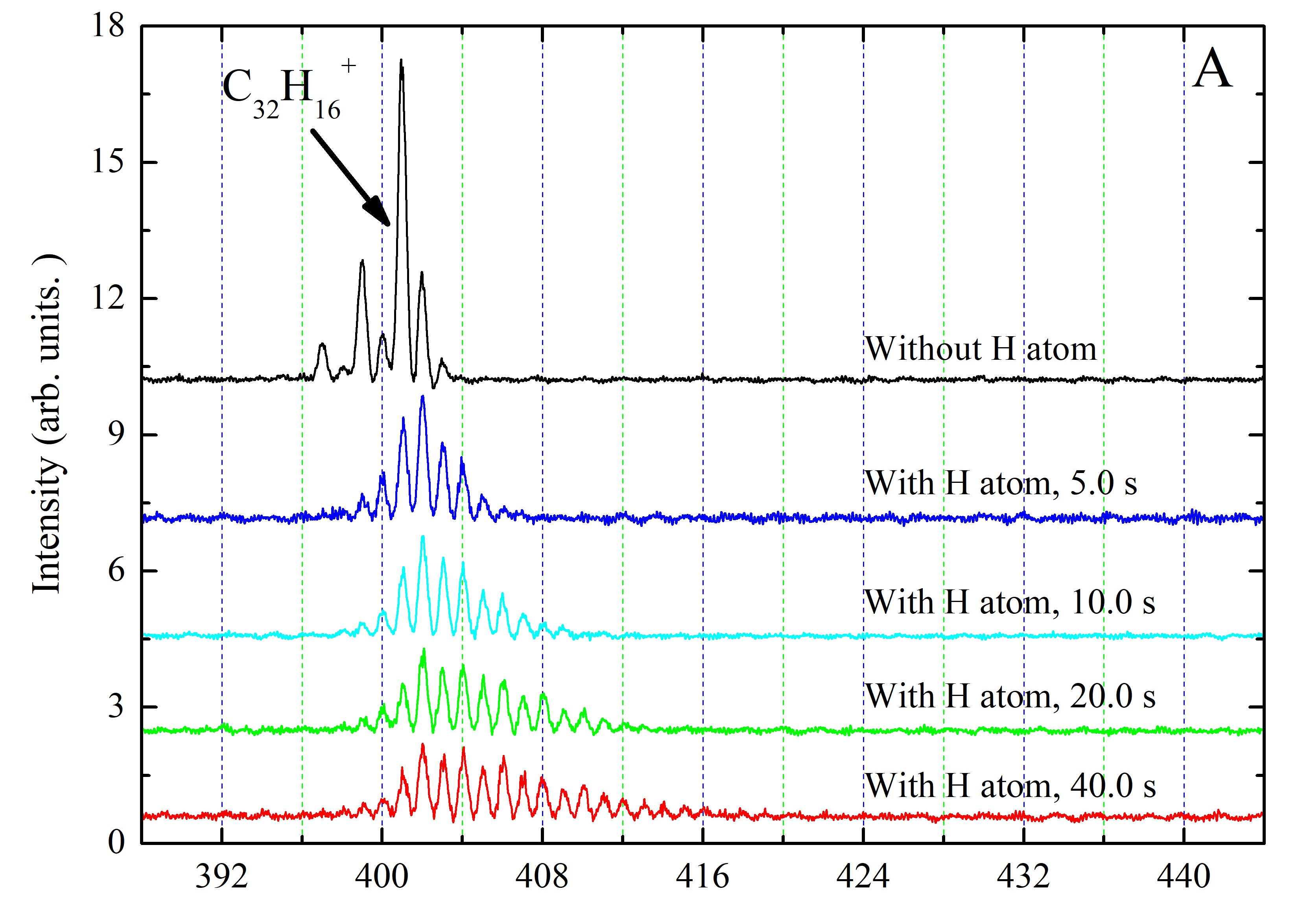}\\
                    \vspace{0.02cm}
                    \includegraphics[width=\linewidth]{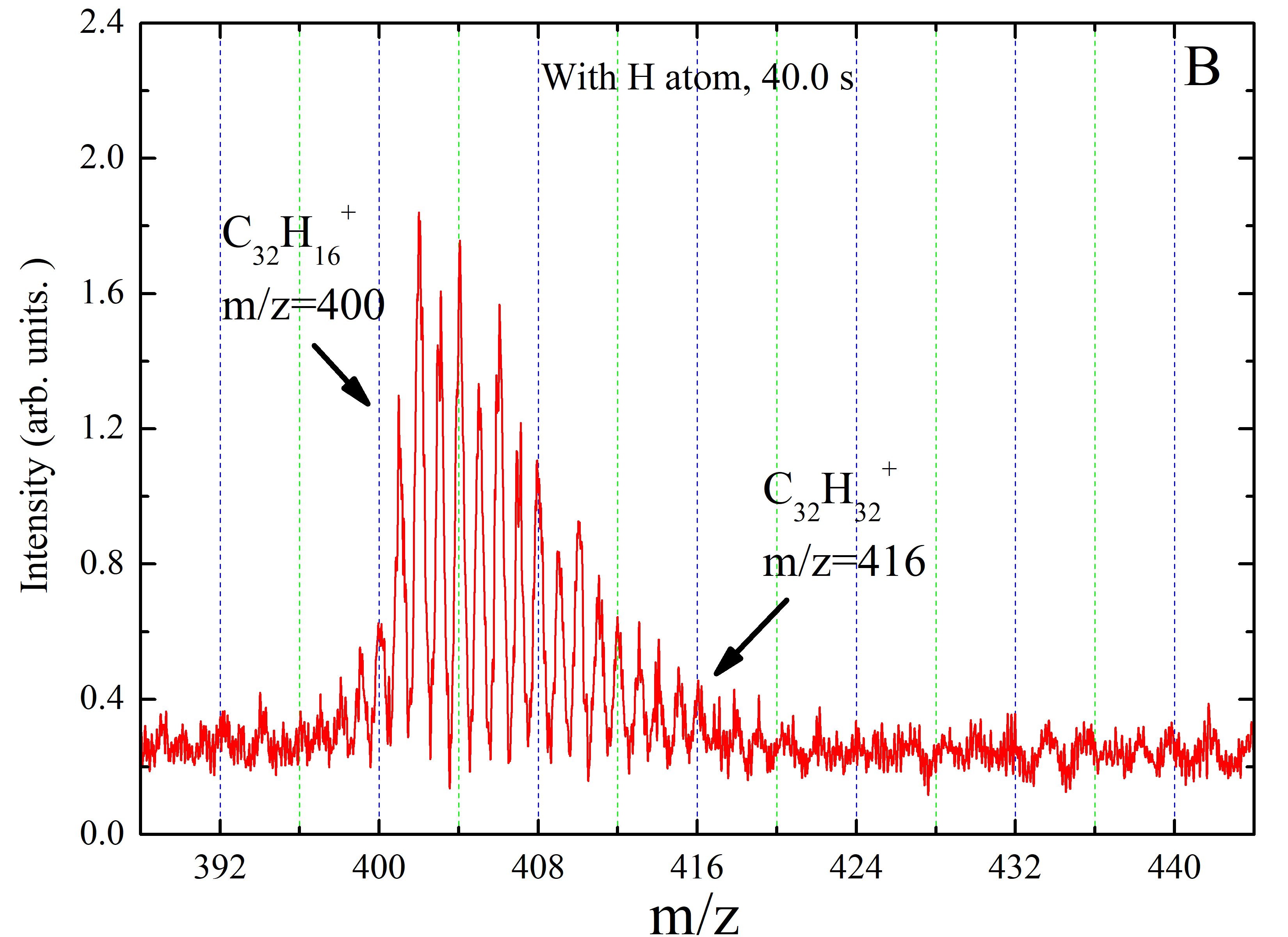}
                    \vspace{0.02cm}
             \end{minipage}%
      }%
      \subfigure{
             \begin{minipage}{0.4\linewidth}
                   \centering
                    \renewcommand\arraystretch{2.28}
                    \setlength{\tabcolsep}{8mm}
                    \begin{tabular}{|c|c|}
                           \hline
                           \multicolumn{2}{|c|}{$\rm{[C_{32}H_{16}]}^+$} \\ \hline
                           \specialrule{0em}{0pt}{0.25pt}
                           \multicolumn{2}{|c|}{
                                  \begin{minipage}[b]{0.55\columnwidth}
                                         \centering
                                        \raisebox{-.5\height}{\includegraphics[width=\linewidth]{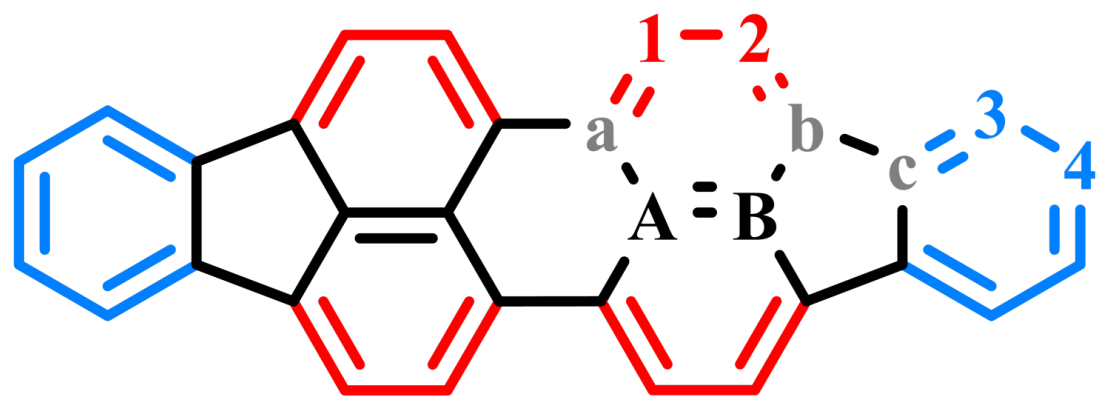}}
                           \end{minipage}} \\ \hline
                           \multicolumn{2}{|c|}{\textbf{Outer carbon sites}} \\ \hline
                           {$\rm{[C_{32}H_{16}+H(1)]}^+$} & -2.02 eV \\ \hline
                           {$\rm{[C_{32}H_{16}+H(2)]}^+$} & -1.45 eV \\ \hline
                           {$\rm{[C_{32}H_{16}+H(3)]}^+$} & -1.78 eV \\ \hline
                           {$\rm{[C_{32}H_{16}+H(4)]}^+$} & -1.94 eV \\ \hline
                           \multicolumn{2}{|c|}{\textbf{Inner carbon sites}} \\ \hline
                           {$\rm{[C_{32}H_{16}+H(a)]}^+$} & -1.12 eV \\ \hline
                           {$\rm{[C_{32}H_{16}+H(b)]}^+$} & -1.69 eV \\ \hline
                           {$\rm{[C_{32}H_{16}+H(c)]}^+$} & -1.48 eV \\ \hline
                           \multicolumn{2}{|c|}{\textbf{Center carbon sites}} \\ \hline
                           {$\rm{[C_{32}H_{16}+H(A)]}^+$} & -0.99 eV \\ \hline
                           {$\rm{[C_{32}H_{16}+H(B)]}^+$} & -1.39 eV \\
                           \hline
                    \end{tabular}
                    \vspace{0.06cm}
             \end{minipage}%
      }%
      \centering
      \caption{Panel (A): evolution of the mass spectrum of the hydrogenated periflanthene cations with increasing H atom exposure time of 5.0, 10.0, 20.0, and 40.0 s; panel (B): zoomed-in mass spectrum (40.0 s), revealing the presence of a highly broadened mass distribution, and the large hydrogenated periflanthene cations $\rm[{C_{32}H_{16+16}}]^+$; right panel: the theoretical calculation results for [C$_{32}$H$_{16}$]$^+$ $+$ H.}
      \vspace{0.02cm}
      \label{fig3}
\end{figure*}

\begin{figure*}
      \subfigure{
             \begin{minipage}[t]{0.5\linewidth}
                    \centering
                    \includegraphics[width=\linewidth]{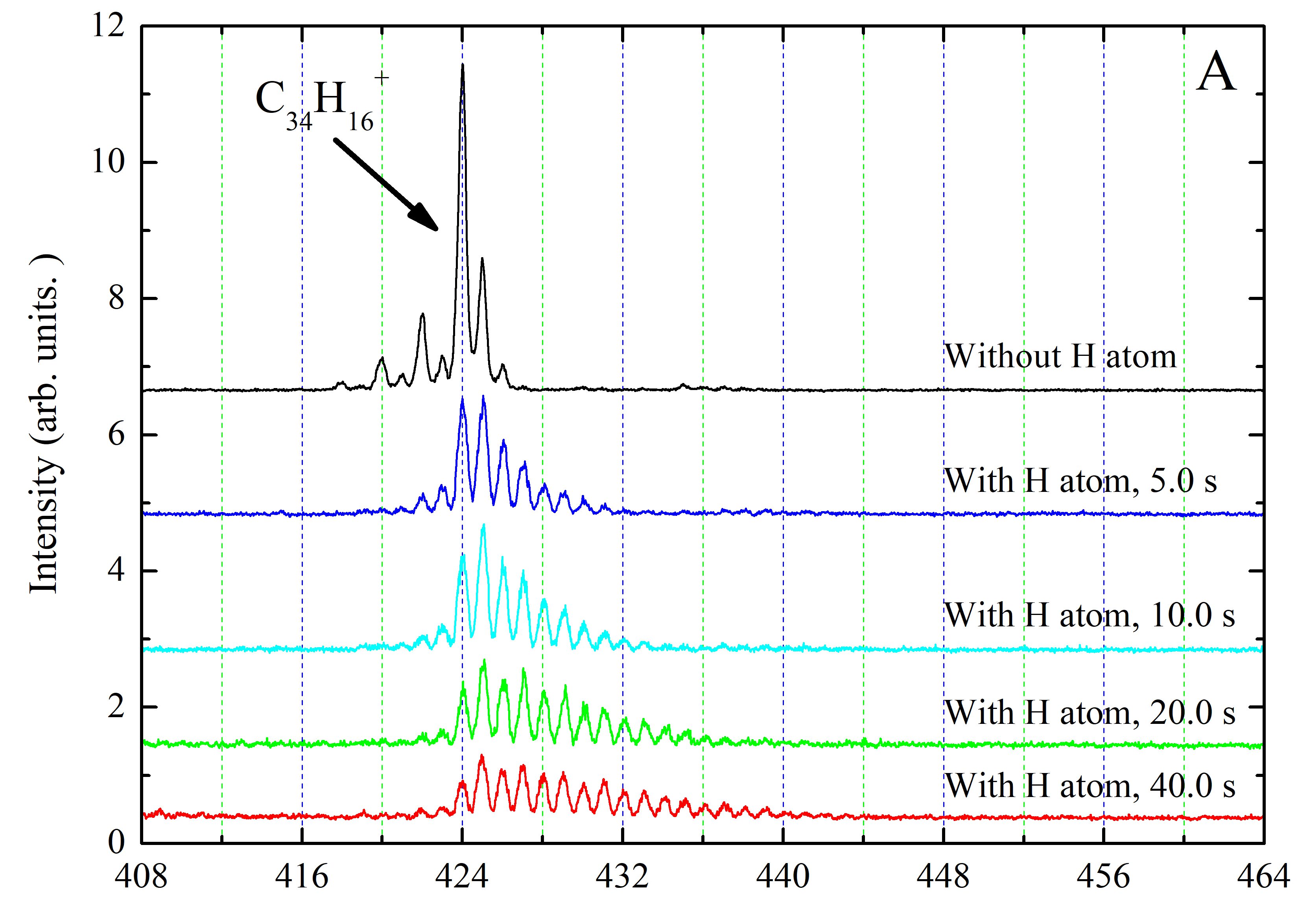}\\
                    \vspace{0.1cm}
                    \includegraphics[width=\linewidth]{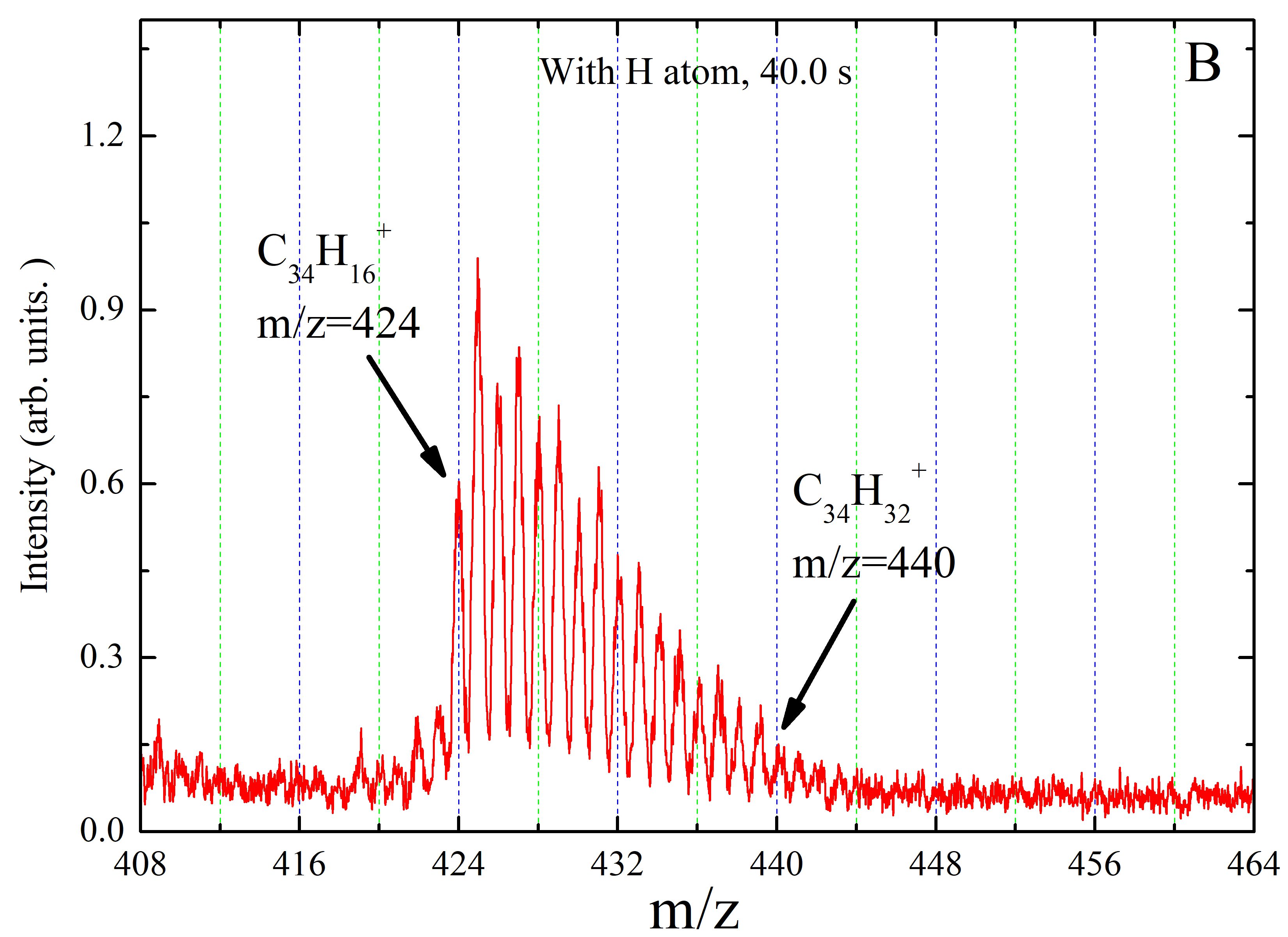}
                    \vspace{0.02cm}
             \end{minipage}%
      }%
      \subfigure{
             \begin{minipage}{0.4\linewidth}
                    \centering
                    \renewcommand\arraystretch{1.42}
                    \setlength{\tabcolsep}{8mm}
                    \begin{tabular}{|c|c|}
                           \hline
                           \multicolumn{2}{|c|}{$\rm{[C_{34}H_{16}]}^+$} \\ \hline
                           \specialrule{0em}{0pt}{0.25pt}
                           \multicolumn{2}{|c|}{
                                  \begin{minipage}[b]{0.32\columnwidth}
                                         \centering
                                        \raisebox{-.5\height}{\includegraphics[width=\linewidth]{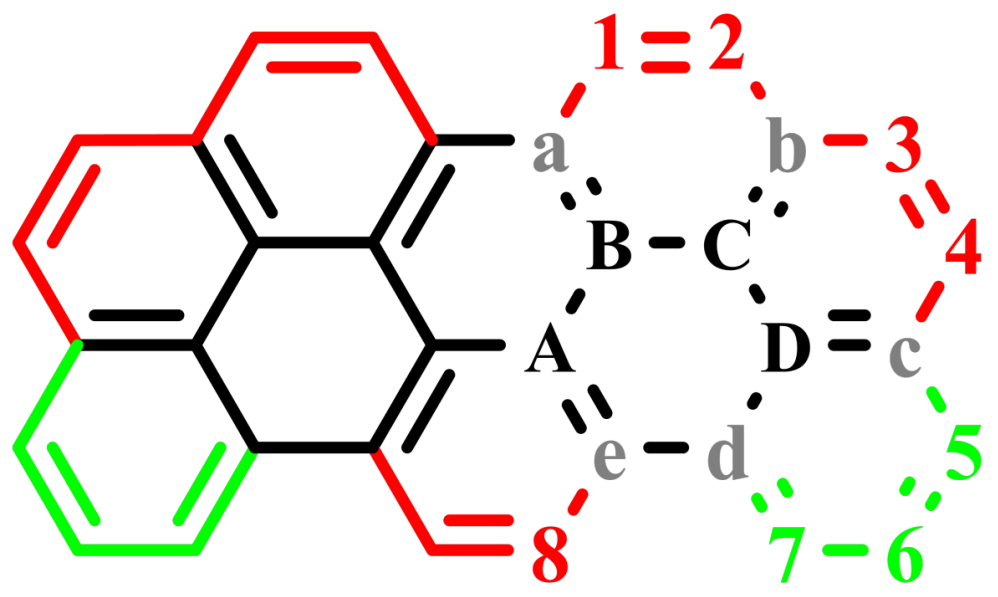}}
                           \end{minipage}} \\ \hline
                           \multicolumn{2}{|c|}{\textbf{Outer carbon sites}} \\ \hline
                           {$\rm{[C_{34}H_{16}+H(1)]}^+$} & -1.78 eV \\ \hline
                           {$\rm{[C_{34}H_{16}+H(2)]}^+$} & -1.70 eV \\ \hline
                           {$\rm{[C_{34}H_{16}+H(3)]}^+$} & -1.43 eV \\ \hline
                           {$\rm{[C_{34}H_{16}+H(4)]}^+$} & -1.83 eV \\ \hline
                           {$\rm{[C_{34}H_{16}+H(5)]}^+$} & -2.10 eV \\ \hline
                           {$\rm{[C_{34}H_{16}+H(6)]}^+$} & -1.31 eV \\ \hline
                           {$\rm{[C_{34}H_{16}+H(7)]}^+$} & -2.10 eV \\ \hline
                           {$\rm{[C_{34}H_{16}+H(8)]}^+$} & -1.73 eV \\ \hline
                           \multicolumn{2}{|c|}{\textbf{Inner carbon sites}} \\ \hline
                           {$\rm{[C_{34}H_{16}+H(a)]}^+$} & -1.25 eV \\ \hline
                           {$\rm{[C_{34}H_{16}+H(b)]}^+$} & -1.22 eV \\ \hline
                           {$\rm{[C_{34}H_{16}+H(c)]}^+$} & -0.77 eV \\ \hline
                           {$\rm{[C_{34}H_{16}+H(d)]}^+$} & -0.92 eV \\ \hline
                           {$\rm{[C_{34}H_{16}+H(e)]}^+$} & -1.31 eV \\ \hline
                           \multicolumn{2}{|c|}{\textbf{Center carbon sites}} \\ \hline
                           {$\rm{[C_{34}H_{16}+H(A)]}^+$} & -1.01 eV \\ \hline
                           {$\rm{[C_{34}H_{16}+H(B)]}^+$} & -1.12 eV \\ \hline
                           {$\rm{[C_{34}H_{16}+H(C)]}^+$} & -0.96 eV \\ \hline
                           {$\rm{[C_{34}H_{16}+H(D)]}^+$} & -0.97 eV \\
                           \hline
                    \end{tabular}
                    \vspace{0.06cm}
             \end{minipage}%
      }%
      \centering
      \caption{Panel (A): evolution of the mass spectrum of the hydrogenated TBP cations with increasing H atom exposure time of 5.0, 10.0, 20.0, and 40.0 s; panel (B): zoomed-in mass spectrum (40.0 s), revealing the presence of a highly broadened mass distribution, and the large hydrogenated TBP cations $\rm[{C_{34}H_{16+16}}]^+$; right panel: the theoretical calculation results for [C$_{34}$H$_{16}$]$^+$ $+$ H.}
      \vspace{0.02cm}
      \label{fig4}
\end{figure*}

\begin{figure*}
      \subfigure{
             \begin{minipage}[t]{0.5\linewidth}
                    \centering
                    \includegraphics[width=\linewidth]{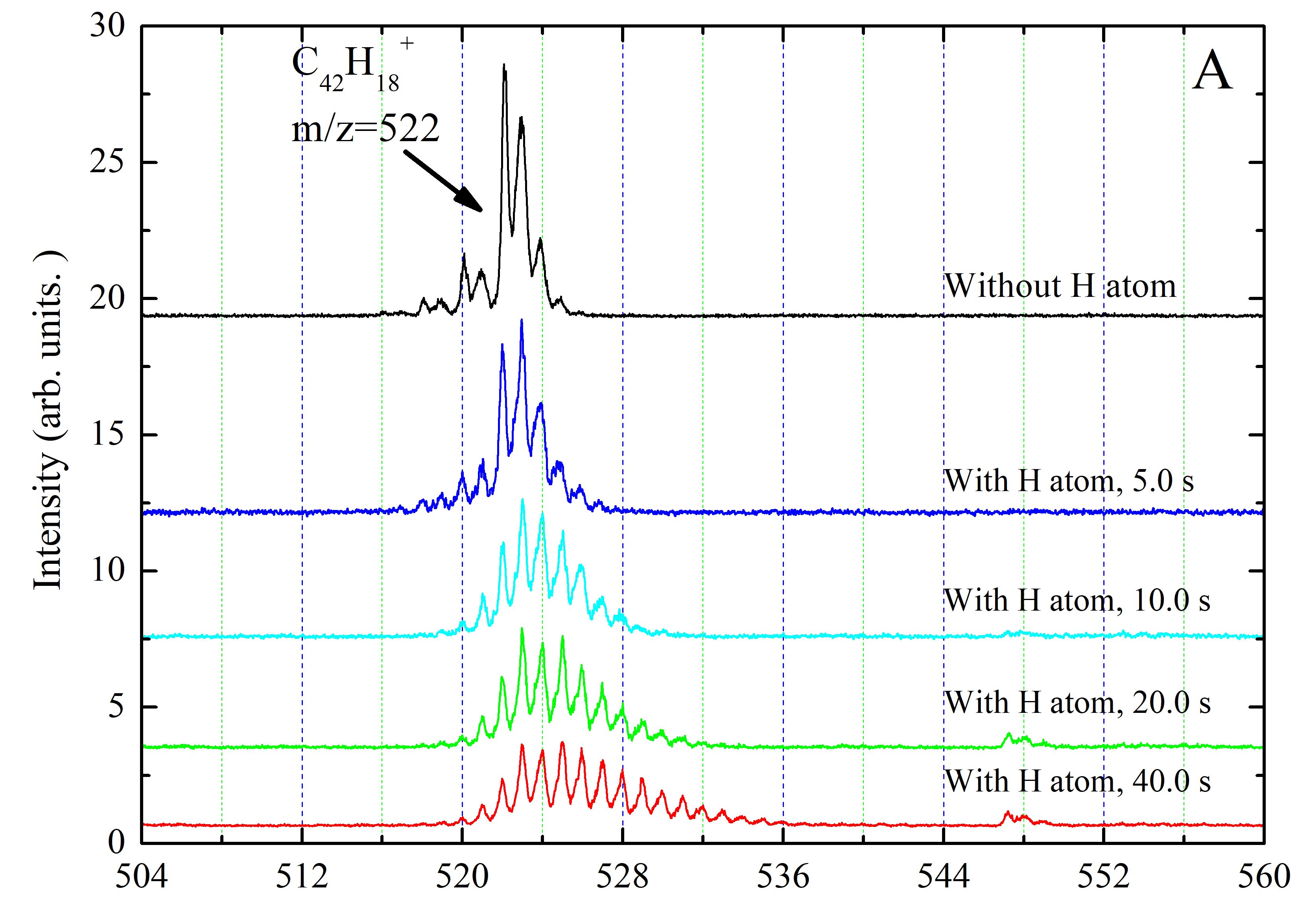}\\
                    \includegraphics[width=\linewidth]{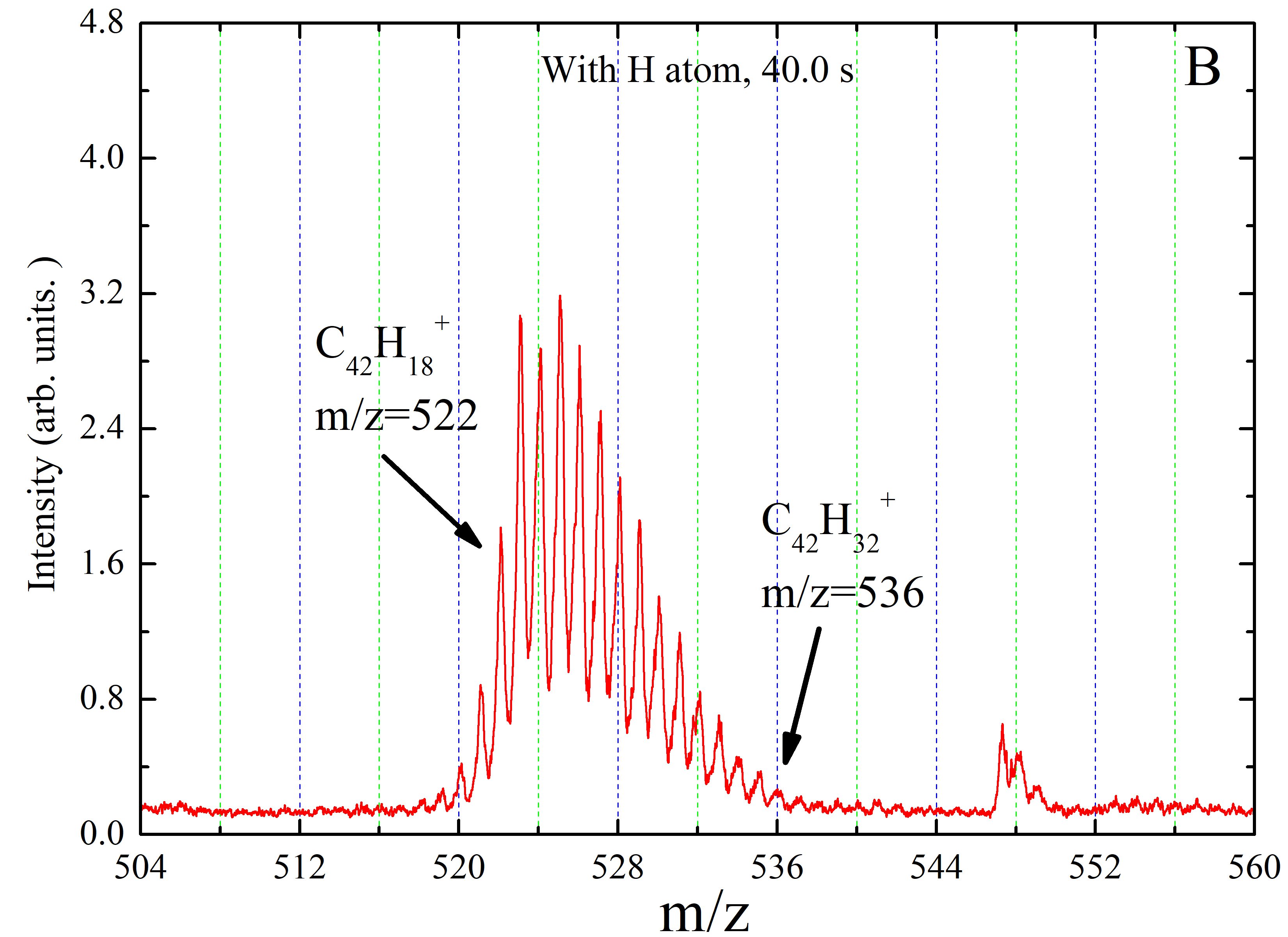}
             \end{minipage}%
      }%
      \subfigure{
             \begin{minipage}{0.4\linewidth}
                    \centering
                    \renewcommand\arraystretch{2.8}
                    \setlength{\tabcolsep}{8mm}
                    \begin{tabular}{|c|c|}
                           \hline
                           \multicolumn{2}{|c|}{$\rm{[C_{42}H_{18}]}^+$} \\ \hline
                           \specialrule{0em}{0pt}{0.25pt}
                           \multicolumn{2}{|c|}{
                                  \begin{minipage}[b]{0.5\columnwidth}
                                         \centering
                                        \raisebox{-.5\height}{\includegraphics[width=\linewidth]{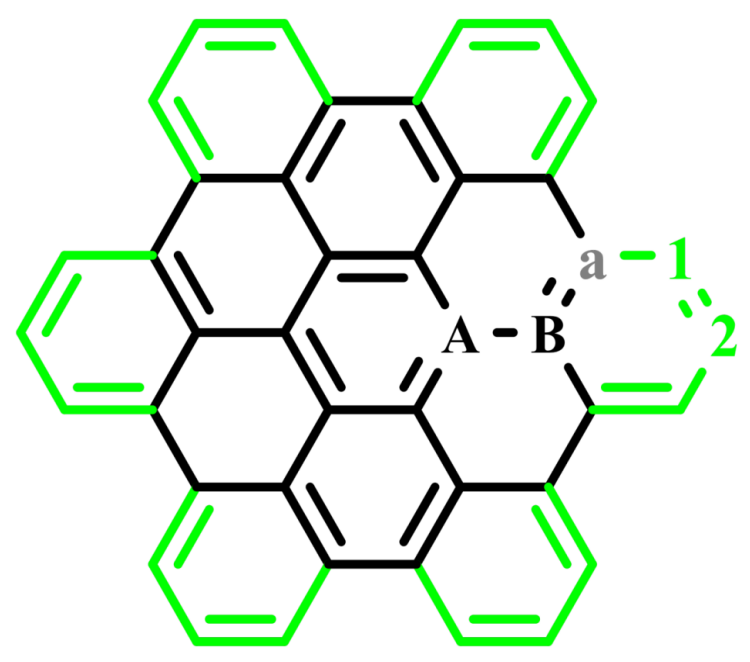}}
                           \end{minipage}} \\ \hline
                           \multicolumn{2}{|c|}{\textbf{Outer carbon sites}} \\ \hline
                           {$\rm{[C_{42}H_{18}+H(1)]}^+$} & -2.05 eV \\ \hline
                           {$\rm{[C_{42}H_{18}+H(2)]}^+$} & -1.62 eV \\ \hline
                           \multicolumn{2}{|c|}{Inner carbon sites} \\ \hline
                           {$\rm{[C_{42}H_{18}+H(a)]}^+$} & -1.25 eV \\ \hline
                           \multicolumn{2}{|c|}{Center carbon sites} \\ \hline
                           {$\rm{[C_{42}H_{18}+H(A)]}^+$} & -1.51 eV \\ \hline
                           {$\rm{[C_{42}H_{18}+H(B)]}^+$} & -1.18 eV \\
                           \hline
                    \end{tabular}
                    \vspace{0.06cm}
             \end{minipage}%
      }%
      \centering
      \caption{Panel (A): evolution of the mass spectrum of the hydrogenated HBC cations with increasing H atom exposure time of 5.0, 10.0, 20.0, and 40.0 s; panel (B): zoomed-in mass spectrum (40.0 s), revealing the presence of a highly broadened mass distribution, and the large hydrogenated HBC cations $\rm[{C_{42}H_{18+14}}]^+$; right panel: the theoretical calculation results for [C$_{42}$H$_{18}$]$^+$ $+$ H.}
      \vspace{0.02cm}
      \label{fig5}
\end{figure*}

\begin{figure*}
      \subfigure{
             \begin{minipage}[t]{0.5\linewidth}
                    \centering
                    \includegraphics[width=\linewidth]{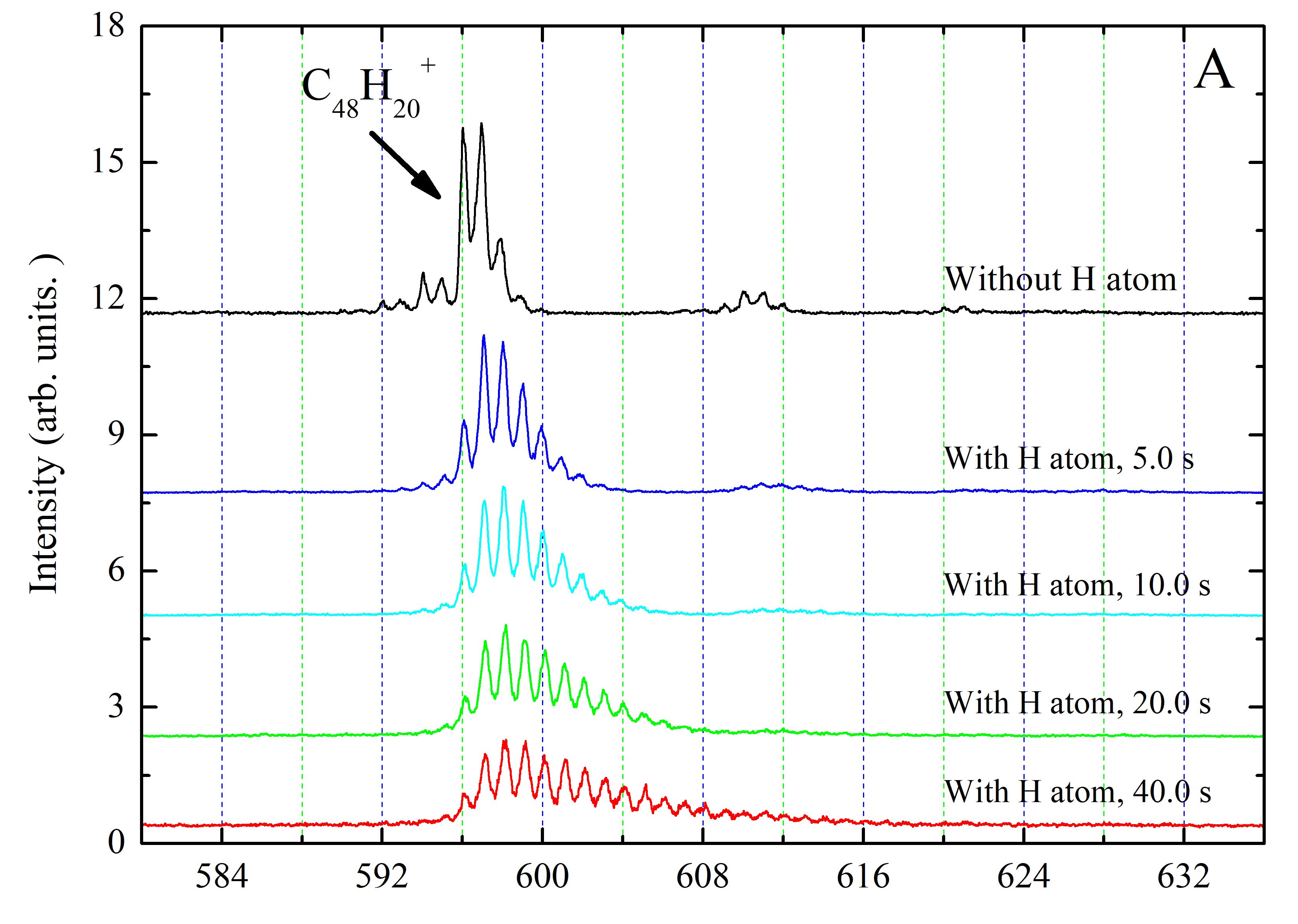}\\
                    \vspace{0.1cm}
                    \includegraphics[width=\linewidth]{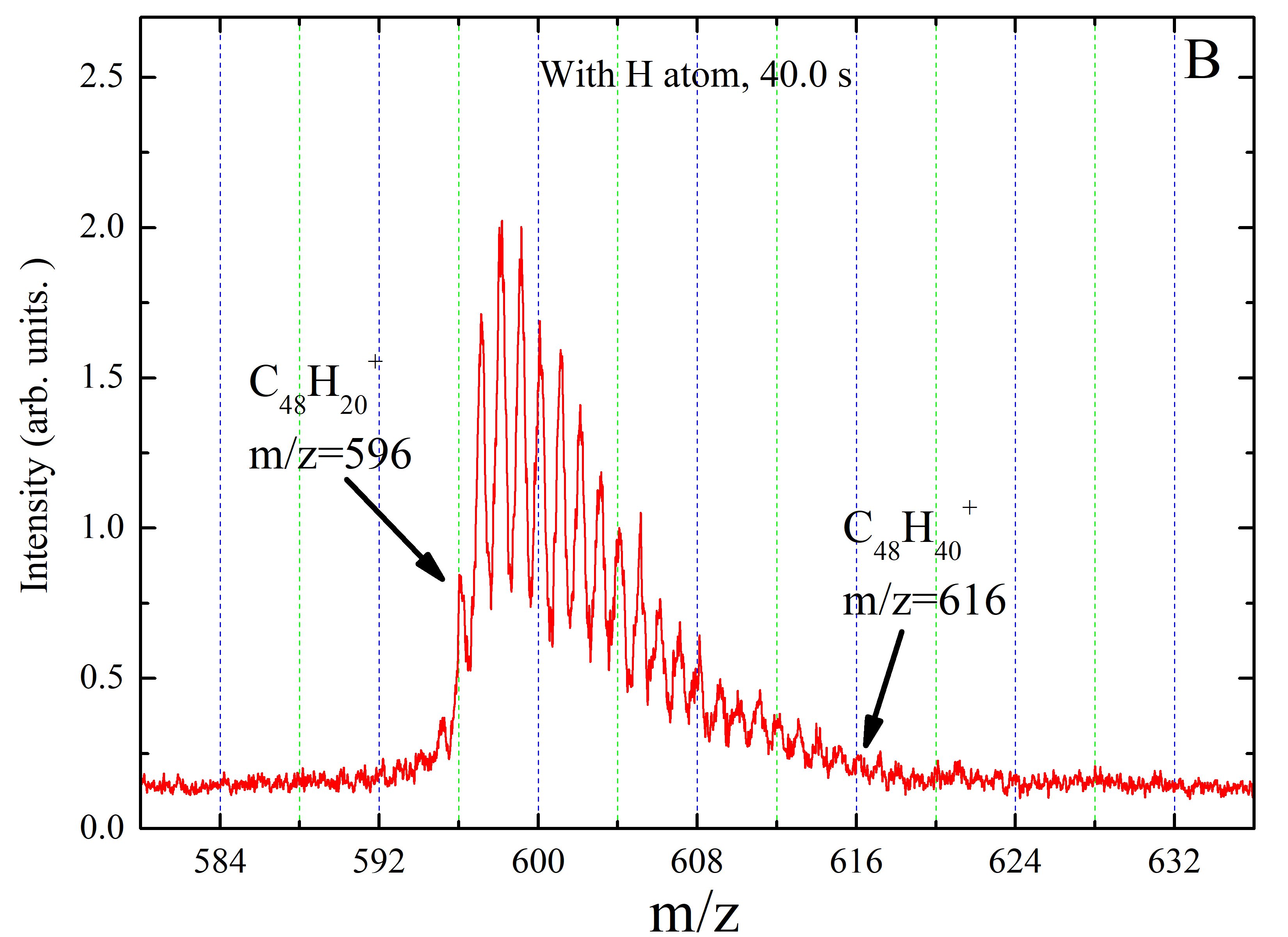}
                    \vspace{0.02cm}
             \end{minipage}%
      }%
      \subfigure{
             \begin{minipage}{0.4\linewidth}
                    \centering
                    \renewcommand\arraystretch{1.6}
                    \setlength{\tabcolsep}{8mm}
                    \begin{tabular}{|c|c|}
                           \hline
                           \multicolumn{2}{|c|}{$\rm{[C_{48}H_{20}]}^+$} \\ \hline
                           \specialrule{0em}{0pt}{0.25pt}
                           \multicolumn{2}{|c|}{
                                  \begin{minipage}[b]{0.48\columnwidth}
                                         \centering
                                        \raisebox{-.5\height}{\includegraphics[width=\linewidth]{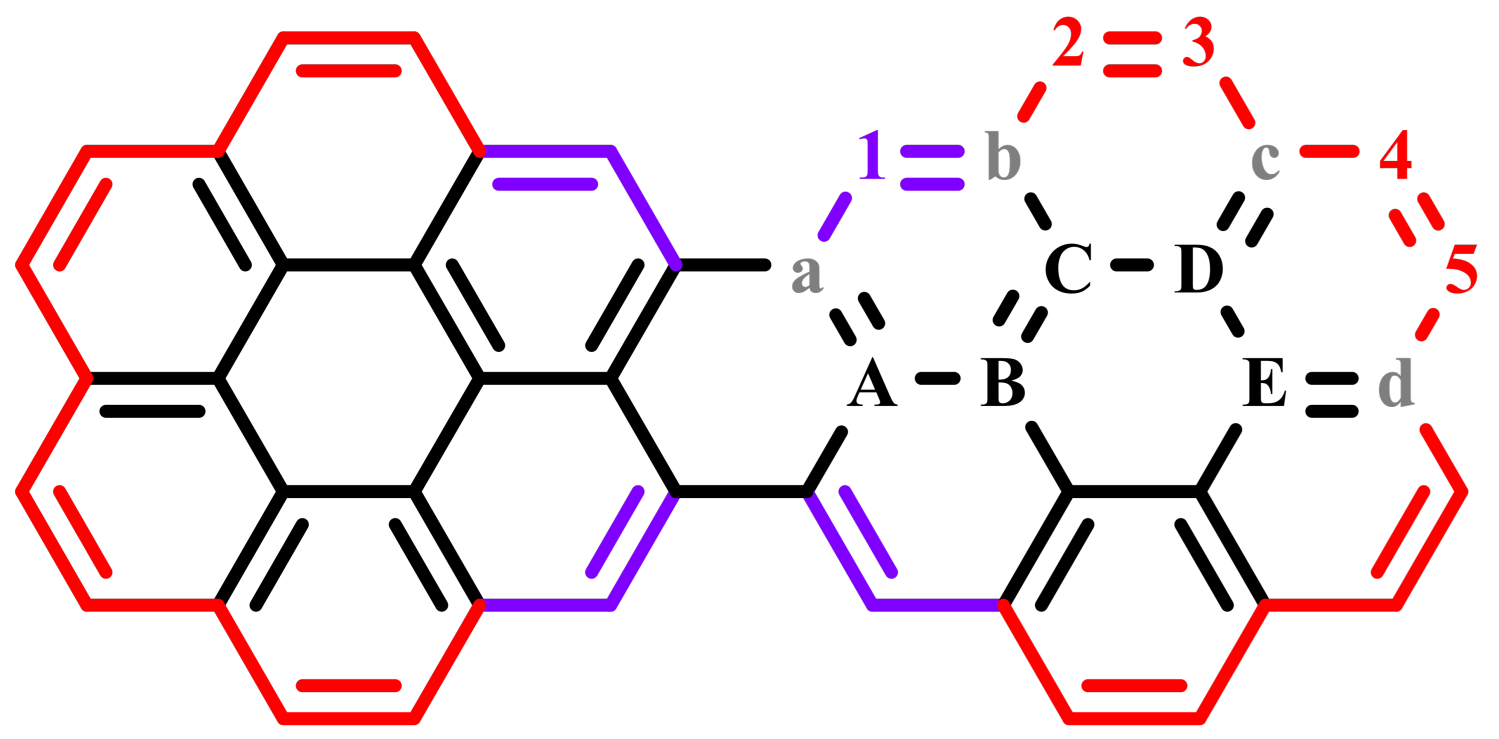}}
                           \end{minipage}} \\ \hline
                           \multicolumn{2}{|c|}{\textbf{Outer carbon sites}} \\ \hline
                           {$\rm{[C_{48}H_{20}+H(1)]}^+$} & -2.08 eV \\ \hline
                           {$\rm{[C_{48}H_{20}+H(2)]}^+$} & -1.74 eV \\ \hline
                           {$\rm{[C_{48}H_{20}+H(3)]}^+$} & -1.47 eV \\ \hline
                           {$\rm{[C_{48}H_{20}+H(4)]}^+$} & -1.46 eV \\ \hline
                           {$\rm{[C_{48}H_{20}+H(5)]}^+$} & -1.77 eV \\ \hline
                           \multicolumn{2}{|c|}{Inner carbon sites} \\ \hline
                           {$\rm{[C_{48}H_{20}+H(a)]}^+$} & -1.04 eV \\ \hline
                           {$\rm{[C_{48}H_{20}+H(b)]}^+$} & -0.80 eV \\ \hline
                           {$\rm{[C_{48}H_{20}+H(c)]}^+$} & -1.16 eV \\ \hline
                           {$\rm{[C_{48}H_{20}+H(d)]}^+$} & -0.86 eV \\ \hline
                           \multicolumn{2}{|c|}{Center carbon sites} \\ \hline
                           {$\rm{[C_{48}H_{20}+H(A)]}^+$} & -1.06 eV \\ \hline
                           {$\rm{[C_{48}H_{20}+H(B)]}^+$} & -0.70 eV \\ \hline
                           {$\rm{[C_{48}H_{20}+H(C)]}^+$} & -1.02 eV \\ \hline
                           {$\rm{[C_{48}H_{20}+H(D)]}^+$} & -0.67 eV \\ \hline
                           {$\rm{[C_{48}H_{20}+H(E)]}^+$} & -0.75 eV \\
                           \hline
                    \end{tabular}
                    \vspace{0.06cm}
             \end{minipage}%
      }%
      \centering
      \caption{Panel (A): evolution of the mass spectrum of the hydrogenated DC cations with increasing H atom exposure time of 5.0, 10.0, 20.0, and 40.0 s; panel (B): zoomed-in mass spectrum (40.0 s), revealing the presence of a highly broadened mass distribution, and the large hydrogenated DC cations $\rm[{C_{48}H_{20+20}}]^+$; right panel: the theoretical calculation results for [C$_{48}$H$_{20}$]$^+$ $+$ H.}
      \vspace{0.02cm}
      \label{fig6}
\end{figure*}

\begin{table*}
      \centering
      \caption{The theoretical calculation results (outer side-edged carbon sites) for [C$_{40}$H$_{18}$]$^+$ $+$ H and [C$_{40}$H$_{19}$]$^+$ $+$ H.The doublet spin multiplicity is considered in all calculations in this table.}
      \label{tab:chap:table2}
      \renewcommand\arraystretch{1.7}%
      \setlength{\tabcolsep}{1.1mm}
      \begin{tabular}[c]{|c|c|c|c|c|c|}
             \hline
             sites diagram & \makecell[c]{edge structure} & add the first H-atoms & E(eV) & add the second H-atoms & E(eV)\\ \hline
             \multirow{20}{*}{\begin{minipage}[b]{0.4\columnwidth}
                           \centering
                           \raisebox{-.5\height}{\includegraphics[width=\linewidth]{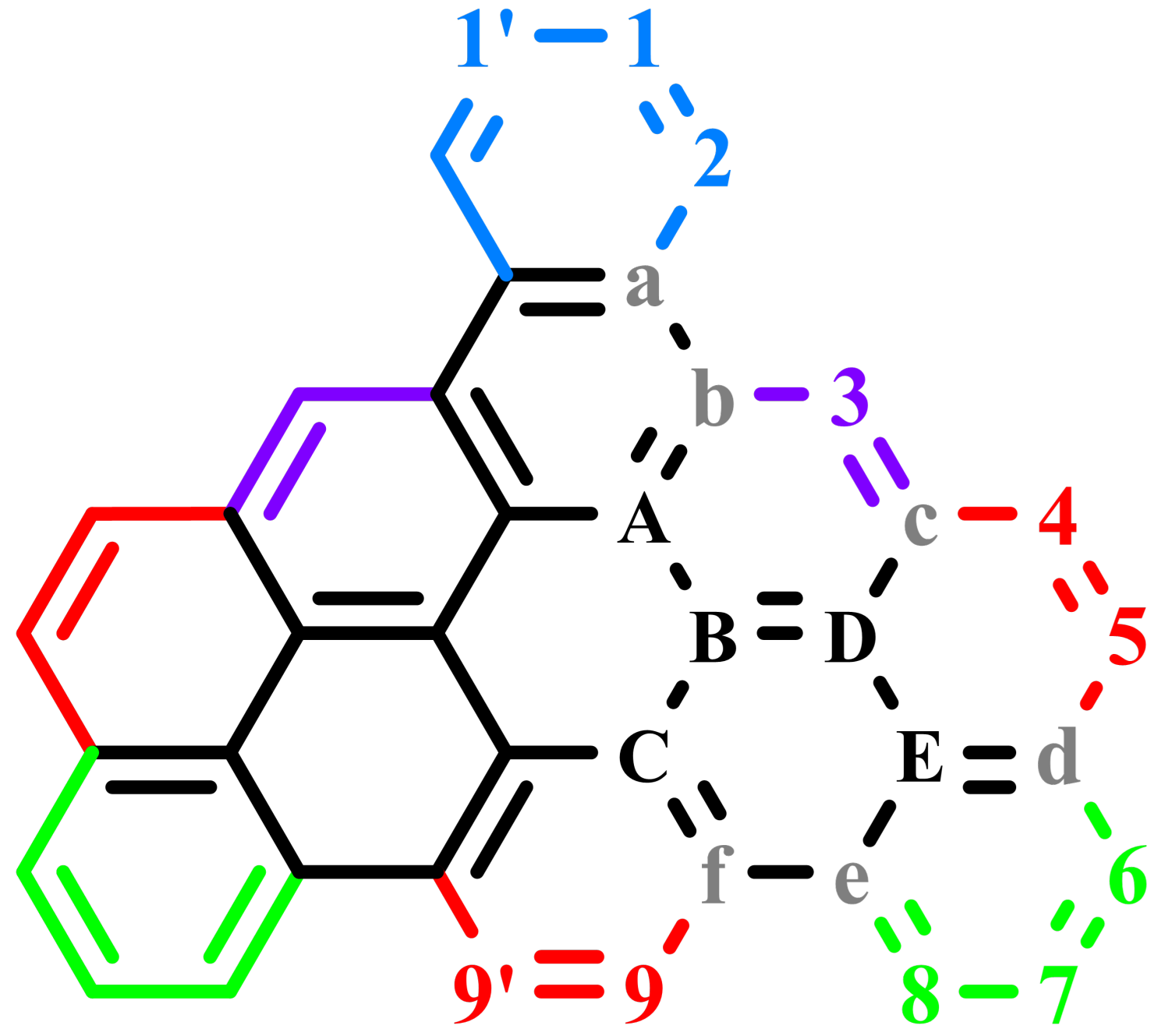}}
             \end{minipage}} & solo & $\rm{[C_{40}H_{18}+H(3)]}^+$ & -2.14 & $\rm{[C_{40}H_{18}]}^+$ + $\rm{H_2}$ & -2.09 \\ \cline{2-6}
             ~ & \multirow{6}{*}{duo} & \multirow{2}{*}{$\rm{[C_{40}H_{18}+H(4)]}^+$} & \multirow{2}{*}{-1.54} & $\rm{[C_{40}H_{18}+H(4,5)]}^+$ & -2.70 \\ \cline{5-6}
             ~ & ~ & ~ & ~ & $\rm{[C_{40}H_{18}]}^+$ + $\rm{H_2}$ & -2.70 \\ \cline{3-6}
             ~ & ~ & \multirow{2}{*}{ $\rm{[C_{40}H_{18}+H(5)]}^+$} & \multirow{2}{*}{-1.79} & $\rm{[C_{40}H_{18}+H(5,4)]}^+$ & -2.44 \\ \cline{5-6}
             ~ & ~ & ~ & ~ & $\rm{[C_{40}H_{18}]}^+$ + $\rm{H_2}$ & -2.44 \\ \cline{3-6}
             ~ & ~ & \multirow{2}{*}{ $\rm{[C_{40}H_{18}+H(9)]}^+$} & \multirow{2}{*}{-1.77} & $\rm{[C_{40}H_{18}+H(9,9')]}^+$ & -2.44 \\ \cline{5-6}
             ~ & ~ & ~ & ~ & $\rm{[C_{40}H_{18}]}^+$ + $\rm{H_2}$ & -2.46 \\ \cline{2-6}
             ~ & \multirow{9}{*}{trio} & \multirow{3}{*}{ $\rm{[C_{40}H_{18}+H(6)]}^+$} & \multirow{3}{*}{-2.10 } & $\rm{[C_{40}H_{18}+H(6,7)]}^+$ & -1.81 \\ \cline{5-6}
             ~ & ~ & ~ & ~ & $\rm{[C_{40}H_{18}+H(6,8)]}^+$ & -1.60 \\ \cline{5-6}
             ~ & ~ & ~ & ~ & $\rm{[C_{40}H_{18}]}^+$ + $\rm{H_2}$ &-2.14 \\ \cline{3-6}
             ~ & ~ & \multirow{3}{*}{ $\rm{[C_{40}H_{18}+H(7)]}^+$} & \multirow{3}{*}{-1.45 } & $\rm{[C_{40}H_{18}+H(7,6)]}^+$ & -2.46 \\ \cline{5-6}
             ~ & ~ & ~ & ~ & $\rm{[C_{40}H_{18}+H(7,8)]}^+$ & -2.47 \\ \cline{5-6}
             ~ & ~ & ~ & ~ & $\rm{[C_{40}H_{18}]}^+$ + $\rm{H_2}$ & -2.78 \\ \cline{3-6}
             ~ & ~ & \multirow{3}{*}{ $\rm{[C_{40}H_{18}+H(8)]}^+$} & \multirow{3}{*}{-2.11 } & $\rm{[C_{40}H_{18}+H(8,6)]}^+$ & -1.80 \\ \cline{5-6}
             ~ & ~ & ~ & ~ & $\rm{[C_{40}H_{18}+H(8,7)]}^+$ & -1.82 \\ \cline{5-6}
             ~ & ~ & ~ & ~ & $\rm{[C_{40}H_{18}]}^+$ + $\rm{H_2}$ & -2.13 \\ \cline{2-6}
             ~ & \multirow{5}{*}{quarto}~ & \multirow{3}{*}{ $\rm{[C_{40}H_{18}+H(1)]}^+$} & \multirow{3}{*}{\makecell[c]{-1.53 } } & $\rm{[C_{40}H_{18}+H(1,1')]}^+$ & -1.98 \\ \cline{5-6}
             ~ & ~ & ~ & ~ & $\rm{[C_{40}H_{18}+H(1,2)]}^+$ & -2.34 \\ \cline{5-6}
             ~ & ~ & ~ & ~ & $\rm{[C_{40}H_{18}]}^+$ + $\rm{H_2}$ & -2.70 \\ \cline{3-6}
             ~ & ~ & \multirow{2}{*}{ $\rm{[C_{40}H_{18}+H(2)]}^+$} & \multirow{2}{*}{ -1.54 } & $\rm{[C_{40}H_{18}+H(2,1)]}^+$ & -2.33 \\ \cline{5-6}
             ~ & ~ & ~ & ~ & $\rm{[C_{40}H_{18}]}^+$ + $\rm{H_2}$ & -2.70 \\ \cline{1-6}
      \end{tabular}
\end{table*}

\subsection{Experimental results and calculation for the hydrogenation of PAH cations}

\subsubsection{Experimental results and calculation for TNPP cations ([C$_{40}$H$_{18}$]$^+$) $+$ H}

In Figure 1, the evolution of the mass spectrum of hydrogenated TNPP cations is presented in the range of m/z=484-540. A series of hydrogenated TNPP cations, [C$_{40}$H$_{18+n}$]$^+$, are observed. As shown in Figure 1(A), the black line is the TNPP cation mass spectrum before hydrogenation. In addition to the main TNPP mass peak, some residual TNPP fragments (peaks due to H loss) were also measured, such as C$_{40}{\rm{H}_{16/17/18/19}}^+$, which are byproducts of the electron impact ionization and fragmentation processes \citep{zhen2014}. This phenomenon is also present in the other mass spectra measured in this paper for the same reason. By varying the H-atoms' exposure time (from 5.0 to 10.0, 20.0, and 40.0 s), the hydrogenation degree of TNPP cations changes. The m/z distribution exhibits the expected shift to higher masses, which indicates that H-atoms are being added to TNPP cations. With H atom exposure time increasing, the average intensity of each peak decreases, but the total intensity of ions trapped in the ion trap is almost unchanged. As shown in Figure 1(B), a zoomed-in mass spectrum (the H atom exposure time is 40.0 s) reveals a highly broadened mass distribution and the presence of newly formed large hydrogenated TNPP cations: m/z=516, [C$_{40}$H$_{18+18}$]$^+$, n=18, which suggest that 18 H-atoms are being added to TNPP cations, the hydrogenated TNPP cations are efficiently formed. In addition, no even-odd hydrogenated mass patterns are observed in the hydrogenation experiments.

Meanwhile, there is no clear mass peaks that can attribute to the PAH fragments (e.g., loss of C$_2$H$_2$) in our obtained mass spectrometry. We suppose that the non-appearance of fragmentation mainly attributes to the superior stability of large PAHs  (listed in Table 1) upon hydrogenation compared to smaller PAH (e. g., C$_{16}$H$_{10}$) \citep{sch20}. And due to that, the large PAHs may play an important role as an ideal catalyst for hydrogen formation without destruction in the ISM \citep{tie13}.

Regarding the hydrogenation pathway of TNPP cations, we believe that these hydrogenated TNPP cations are formed through ion-molecule reaction pathways, i.e., [C$_{40}$H$_{18}$]$^+$ $+$ H-atoms. The reaction process between TNPP cations and H-atoms occurs through sequential steps and repeatedly adds H-atoms. In addition, from previous studies \citep{pet95,lep01,caz16,zhang23}, the competition between hydrogenation and dehydrogenation has been observed in the ion-molecule reaction pathways, i.e., H$_2$ is formed as a secondary product through the reaction pathway as follows:
\begin{small}
      \begin{eqnarray*}
             [\rm {C_{40}H_{18}}]^+ \stackrel{\rm H}{\longrightarrow} [\rm {C_{40}H_{18+n}}]^+ + m*H_2
      \end{eqnarray*}
\end{small}
As shown in Figure 1, TNPP (C$_{40}$H$_{18}$) has 18 outer (solo, duo, trio, and quarto side-edged structures), 12 inner, and 10 center carbon sites, and we identify each carbon skeleton structure's effect on the hydrogenation processes. The carbon reaction sites in symmetrical positions have similar reaction activity. According to that, 20 reaction pathways are considered and obtained in the theoretical calculations. The reaction sites are highlighted with different colors and labeled: solo (purple, 3), duo (red, 4, 5, and 9), trio (green, 6, 7, and 8), and quarto (blue, 1, and 2), inner (a, b, c, d, e, and f) and center (A, B, C, D, and E).

The binding energies of the mono-hydrogenated TNPP cations obtained are given in the right panel of Figure 1. These 20 mono-hydrogenated TNPP cations are divided into three groups: outer carbon edge, inner carbon edge, and center carbon edge.

The first group is for the outer carbon edge, and we can see that all the reaction pathways are exothermic. For quarto side-edged carbon sites, with -1.53 and -1.54 eV for [C$_{40}$H$_{18}$ + H(1/2)]$^+$; For solo side-edged carbon sites, with -2.14 eV for [C$_{40}$H$_{18}$ + H(3)]$^+$; For duo side-edged carbon sites, with -1.54, -1.79, and -1.77 eV for [C$_{40}$H$_{18}$ + H(4/5/9)]$^+$; For trio side-edged carbon sites, with -2.10, -1.45, and -2.11 eV for [C$_{40}$H$_{18}$ + H(6/7/8)]$^+$; The second group is for the inner carbon edge, we can see all the reaction pathways are exothermic. With -1.09, -1.30, -1.16, -0.93, -1.03, and -1.37 eV for [C$_{40}$H$_{18}$ + H(a/b/c/d/e/f)]$^+$. The third group is for the center carbon edge, and we can see that all the reaction pathways are exothermic. With -1.31, -1.19, -1.03, -1.22, and -0.98 eV for [C$_{40}$H$_{18}$ + H(A/B/C/D/E)]$^+$, respectively.

The exothermic energy for the formation of [C$_{40}$H$_{19}$]$^+$, in the range of 0.93-2.14 eV, is relatively higher and can stabilize the whole molecules. The ion-molecule reactions between TNPP cations and H-atoms readily occur, and we propose that [C$_{40}$H$_{19}$]$^+$ formed in the lab are a mixture with different possible isomers.

\subsubsection{Experimental results and calculation for ovalene cations ([C$_{32}$H$_{14}$]$^+$) $+$ H}

In Figure 2, the evolution of the mass spectrum of hydrogenated ovalene cations is presented in the range of m/z=384-440. A series of hydrogenated ovalene cations, [C$_{32}$H$_{14+n}$]$^+$, are observed. As shown in Figure 2(A), by varying the H-atoms' exposure time (from 5.0 to 10.0, 20.0, and 40.0 s), the hydrogenation degree of ovalene cations changes. The m/z distribution exhibits the expected shift to higher masses, which indicates that H-atoms are being added to ovalene cations. With H atom exposure time increasing, the average intensity of each peak decreases, but the total intensity of ions trapped in the ion trap is almost unchanged. As shown in Figure 2(B), a zoomed-in mass spectrum (the H-atoms exposure time is 40.0 s) reveals a highly broadened mass distribution and the presence of newly formed large hydrogenated ovalene cations: m/z=412, [C$_{32}$H$_{14+14}$]$^+$, n=14, which suggest that 14 H-atoms are being added to ovalene cations, the hydrogenated ovalene cations are efficiently formed.

Regarding the hydrogenation pathway of ovalene cations, similar to the TNPP cations and previous work \citep{caz16,zhang23}, we believe that these hydrogenated ovalene cations are formed through ion-molecule reaction pathways, i.e., [C$_{32}$H$_{14}$]$^+$ $+$ H-atoms. The reaction process between ovalene cations and H-atoms occurs through sequential steps and repeatedly adds H-atoms. In addition, H$_2$ is formed as a secondary product through the reaction pathway as follows:
\begin{small}
      \begin{eqnarray*}
            [\rm {C_{32}H_{14}}]^+ \stackrel{\rm H}{\longrightarrow} [\rm {C_{32}H_{14+n}}]^+ + m*H_2
      \end{eqnarray*}
\end{small}
As shown in Figure 2, ovalene ($\rm{C_{32}H_{14}}$) has 14 outer (solo and duo side-edged structures), 8 inner, and 10 center carbon sites, and we will identify each carbon skeleton structure's effect on the hydrogenation processes. The carbon reaction sites in symmetrical positions have similar reaction activity. According to that, 9 reaction pathways are considered and obtained in the theoretical calculations. The reaction sites are highlighted with different colors and labeled solo (purple, 1), duo (red, 2, 3, and 4), inner (a and b), and center (A, B, and C).

The binding energies of the mono-hydrogenated ovalene cations are given in the right panel of Figure 2. These 9 mono-hydrogenated ovalene cations are divided into three hydrogenation groups: outer carbon edge, inner carbon edge, and center carbon edge. The first group is for the outer carbon edge, and we can see that all the reaction pathways are exothermic. For solo side-edged carbon sites, with -2.31 eV for [C$_{32}$H$_{14}$ + H(1)]$^+$; For duo side-edged carbon sites, with -1.84, -1.59, and -1.82 eV for [C$_{32}$H$_{14}$ + H(2/3/4)]$^+$; The second group is for the inner carbon edge, we can see all the reaction pathways are exothermic. With -0.94, and -1.23 eV for [C$_{32}$H$_{14}$ + H(a/b)]$^+$, respectively. The third group is for the center carbon edge, and we can see that all the reaction pathways are exothermic. With -0.98, -1.14, and -0.96 eV for [C$_{32}$H$_{14}$ + H(A/B/C)]$^+$, respectively.

The exothermic energy for the formation of [C$_{32}$H$_{15}$]$^+$, in the range of 0.94-2.31 eV, is relatively higher and can stabilize the whole molecules. The ion-molecule reactions between ovalene cations and H-atoms readily occur, and so we propose that [C$_{32}$H$_{15}$]$^+$ formed in the lab are a mixture with different possible isomers.

\subsubsection{Experimental results and calculation for periflanthene cations ([C$_{32}$H$_{16}$]$^+$) $+$ H}

In Figure 3, the evolution of the mass spectrum of hydrogenated periflanthene cations is presented in the range of m/z=388-444. A series of hydrogenated periflanthene cations, [C$_{32}$H$_{16+n}$]$^+$, are observed. As shown in Figure 3(A), by varying the H-atoms' exposure time (from 5.0 to 10.0, 20.0, and 40.0 s), the hydrogenation degree of periflanthene cations changes. The m/z distribution exhibits the expected shift to higher masses, which indicates that H-atoms are being added to periflanthene cations. As shown in Figure 3(B), a zoomed-in mass spectrum (the H atom exposure time is 40.0 s) reveals a highly broadened mass distribution and the presence of newly formed large hydrogenated periflanthene cations: m/z=416, [C$_{32}$H$_{16+16}$]$^+$, n=16, which suggest that 16 H-atoms are being added to periflanthene cations, the hydrogenated periflanthene cations are efficiently formed.

Regarding the hydrogenation pathway of periflanthene cations, we believe that these hydrogenated periflanthene cations are formed through ion-molecule reaction pathways, i.e., [C$_{32}$H$_{16}$]$^+$ $+$ H-atoms. The reaction process between periflanthene cations and H-atoms occurs through sequential steps and repeatedly adds H-atoms. In addition, H$_2$ is formed as a secondary product through the reaction pathway as follows:
\begin{small}
      \begin{eqnarray*}
             [\rm {C_{32}H_{16}}]^+ \stackrel{\rm H}{\longrightarrow} [\rm {C_{32}H_{16+n}}]^+ + m*H_2
      \end{eqnarray*}
\end{small}
As shown in Figure 3, periflanthene (C$_{32}$H$_{16}$) has 16 outer (duo and quarto side-edged structures), 12 inner, and 4 center carbon sites, and we identify each carbon skeleton structure's effect on the hydrogenation processes. The reaction sites in symmetrical positions have similar reaction activity. According to that, 9 reaction pathways are considered and obtained in the theoretical calculations. The reaction sites are highlighted with different colors and labeled duo (red, 1, and 2), quarto (blue, 3, and 4), inner (a, b, and c), and center (A and B).

The binding energies of the mono-hydrogenated periflanthene cations are given in the right panel of Figure 3. These 9 mono-hydrogenated periflanthene cations are divided into three hydrogenation groups: outer edge, inner edge, and center edge. The first group is for the outer carbon edge, and we can see that all the reaction pathways are exothermic. For duo side-edged carbon sites, with -2.02 and -1.45 eV for [C$_{32}$H$_{16}$ + H(1/2)]$^+$; For quarto side-edged carbon sites, with -1.78 and -1.94 eV for [C$_{32}$H$_{16}$ + H(3/4)]$^+$; The second group is for the inner carbon edge, we can see all the reaction pathways are exothermic. With -1.12, -1.69, and -1.48 eV for [C$_{32}$H$_{16}$ + H(a/b/c)]$^+$. The third group is for the center carbon edge, and we can see that all the reaction pathways are exothermic. With -0.99 and -1.39 eV for [C$_{32}$H$_{16}$ + H(A/B)]$^+$, respectively.

The exothermic energy for the formation of [C$_{32}$H$_{17}$]$^+$, in the range of 0.99-2.02 eV, is relatively higher and can stabilize the whole molecules. The ion-molecule reactions between ovalene cations and H-atoms readily occur, and so we propose that [C$_{32}$H$_{17}$]$^+$ formed in the lab are a mixture with different possible isomers.

\subsubsection{Experimental results and calculation for TBP cations ([C$_{34}$H$_{16}$]$^+$) $+$ H}

In Figure 4, the evolution of the mass spectrum of hydrogenated TBP cations is presented in the range of m/z=408-464. A series of hydrogenated TBP cations, [C$_{34}$H$_{16+n}$]$^+$, are observed. As shown in Figure 4(A), by varying the H-atoms' exposure time (from 5.0 to 10.0, 20.0, and 40.0 s), the hydrogenation degree of TBP cations changes. The m/z distribution exhibits the expected shift to higher masses, which indicates that H-atoms are being added to TBP cations. As shown in Figure 4(B), a zoomed-in mass spectrum (the H-atoms exposure time is 40.0 s) reveals a highly broadened mass distribution and the presence of newly formed large hydrogenated TBP cations: m/z=440, [C$_{34}$H$_{16+16}$]$^+$, n=16, which suggest that 16 H-atoms are being added to TBP cations, the hydrogenated TBP cations are efficiently formed.

Regarding the hydrogenation pathway of TBP cations, we believe that these hydrogenated TBP cations are formed through ion-molecule reaction pathways, i.e., [C$_{34}$H$_{16}$]$^+$ $+$ H-atoms. The reaction process between TBP cations and H-atoms occurs through sequential steps and repeatedly adds H-atoms. In addition, H$_2$ is formed as a secondary product through the reaction pathway as follows:
\begin{small}
      \begin{eqnarray*}
             [\rm {C_{34}H_{16}}]^+ \stackrel{\rm H}{\longrightarrow} [\rm {C_{34}H_{16+n}}]^+ + m*H_2
      \end{eqnarray*}
\end{small}
As shown in Figure 4, TBP (C$_{34}$H$_{16}$) has 16 outer (duo and trio side-edged structures), 10 inner, and 8 center carbon sites, and we identify each carbon skeleton structure's effect on the hydrogenation processes. The reaction sites in symmetrical positions have similar reaction activity. According to that, 17 reaction pathways are considered and obtained. The reaction sites are highlighted with different colors and labeled: duo (red, 1, 2, 3, 4, and 8), trio (green, 5, 6, and 7), inner (a, b, c, d, and e) and center (A, B, C, and D).

The binding energies of the mono-hydrogenated TBP cations are given in the right panel of Figure 4. These 17 mono-hydrogenated TBP cations are divided into three hydrogenation groups: outer carbon edge, inner carbon edge, and center carbon edge. The first group is for the outer carbon edge, and we can see that all the reaction pathways are exothermic. For duo side-edged carbon sites, with -1.78, -1.70, -1.43, -1.83, and -1.73 eV for [C$_{34}$H$_{16}$ + H(1/2/3/4/8)]$^+$; For trio side-edged carbon sites, with -2.10, -1.31, and -2.10 eV for [C$_{34}$H$_{16}$ + H(5/6/7)]$^+$; The second group is for the inner carbon edge, we can see all the reaction pathways are exothermic. With -1.25, -1.22, -0.77, -0.92, and -1.31 eV for [C$_{34}$H$_{16}$ + H(a/b/c/d/e)]$^+$. The third group is for the center carbon edge, and we can see that all the reaction pathways are exothermic. With -1.01, -1.12, -0.96, and -0.97 eV for [C$_{34}$H$_{16}$ + H(A/B/C/D)]$^+$, respectively.

The exothermic energy for the formation of [C$_{34}$H$_{17}$]$^+$, in the range of 0.77-2.10 eV, is relatively higher and can stabilize the whole molecules. The ion-molecule reactions between ovalene cations and H-atoms readily occur, and so we propose that [C$_{34}$H$_{17}$]$^+$ formed in the lab are a mixture with different possible isomers.

\subsubsection{Experimental results and calculation for HBC cations ([C$_{42}$H$_{18}$]$^+$) $+$ H}

In Figure 5, the evolution of the mass spectrum of hydrogenated HBC cations is presented in the range of m/z=504-560. A series of hydrogenated HBC cations, [C$_{42}$H$_{18+n}$]$^+$, are observed. As shown in Figure 5(A), by varying the H-atoms' exposure time (from 5.0 to 10.0, 20.0, and 40.0 s), the hydrogenation degree of HBC cations changes. The m/z distribution exhibits the expected shift to higher masses, which indicates that H-atoms are being added to HBC cations. As shown in figure 5(B), a zoomed-in mass spectrum (the H-atoms exposure time is 40.0 s) reveals a highly broadened mass distribution and the presence of newly formed large hydrogenated HBC cations: m/z=536, [C$_{42}$H$_{18+14}$]$^+$, n=14, which suggest that 14 H-atoms are being added to HBC cations, the hydrogenated HBC cations are efficiently formed.

Regarding the hydrogenation pathway of HBC cations, we believe that these hydrogenated HBC cations are formed through ion-molecule reaction pathways, i.e., [C$_{42}$H$_{18}$]$^+$ $+$ H-atoms. The reaction process between HBC cations and H-atoms occurs through sequential steps and repeatedly adds H-atoms. In addition, H$_2$ is formed as a secondary product through the reaction pathway as follows:
\begin{small}
      \begin{eqnarray*}
             [\rm {C_{42}H_{18}}]^+ \stackrel{\rm H}{\longrightarrow} [\rm {C_{42}H_{18+n}}]^+ + m*H_2
      \end{eqnarray*}
\end{small}
As shown in Figure 5, HBC (C$_{42}$H$_{18}$) has 18 outer (trio side-edged structures), 12 inner, and 12 center carbon sites, and we identify each carbon skeleton structure's effect on the hydrogenation processes. The reaction sites in symmetrical positions have similar reaction activity. According to that, 5 reaction pathways are considered and obtained. The reaction sites are highlighted with different colors and labeled trio (green, 1 and 2), inner (a), and center (A and B).

The binding energies of the mono-hydrogenated HBC cations are given in the right panel of Figure 5. These 5 mono-hydrogenated HBC cations are divided into three hydrogenation groups: outer edge, inner edge, and center edge. The first group is for the outer carbon edge, and we can see that all the reaction pathways are exothermic. For trio side-edged carbon sites, with -2.05 and -1.62 eV for [C$_{42}$H$_{18}$ + H(1/2)]$^+$; The second group is for the inner carbon edge, we can see the reaction pathway is exothermic. With -1.25 eV for [C$_{42}$H$_{18}$ + H(a)]$^+$. The third group is for the center carbon edge, and we can see that all the reaction pathways are exothermic. With -1.51 and -1.18 eV for [C$_{42}$H$_{18}$ + H(A/B)]$^+$, respectively.

The exothermic energy for the formation of [C$_{42}$H$_{19}$]$^+$, in the range of 1.18-2.05 eV, is relatively higher and can stabilize the whole molecules. The ion-molecule reactions between ovalene cations and H-atoms readily occur, and so we propose that [C$_{42}$H$_{19}$]$^+$ formed in the lab are a mixture with different possible isomers. In addition, the gas-phase hydrogen/deuterium exchange on HBC cations is studied \citep{zhang22}. The exothermic energy for deuteration and dedeuteration reaction pathway is relatively high, and the value of exothermic energy is consistent with the results that obtained in here.

\subsubsection{Experimental results and calculation for DC cations ([C$_{48}$H$_{20}$]$^+$) $+$ H}

In Figure 6, the evolution of the mass spectrum of hydrogenated DC cations is presented in the range of m/z=580-636. A series of hydrogenated DC cations, [C$_{48}$H$_{20+n}$]$^+$, are observed. As shown in Figure 6(A), by varying the H-atoms' exposure time (from 5.0 to 10.0, 20.0, and 40.0 s), the hydrogenation degree of DC cations changes. The m/z distribution exhibits the expected shift to higher masses, which indicates that H-atoms are being added to DC cations. As shown in Figure 6(B), a zoomed-in mass spectrum (the H-atoms exposure time is 40.0 s) reveals a highly broadened mass distribution and the presence of newly formed large hydrogenated DC cations: m/z=616, [C$_{48}$H$_{20+20}$]$^+$, n=20, which suggest that 20 H-atoms are being added to DC cations, the hydrogenated DC cations are efficiently formed.

Regarding the hydrogenation pathway of DC cations, we believe that these hydrogenated DC cations are formed through ion-molecule reaction pathways, i.e., [C$_{48}$H$_{20}$]$^+$ $+$ H-atoms. The reaction process between DC cations and H-atoms occurs through sequential steps and repeatedly adds H-atoms. In addition, H$_2$ is formed as a secondary product through the reaction pathway as follows:
\begin{small}
      \begin{eqnarray*}
             [\rm {C_{48}H_{20}}]^+ \stackrel{\rm H}{\longrightarrow} [\rm {C_{48}H_{20+n}}]^+ + m*H_2
      \end{eqnarray*}
\end{small}
As shown in Figure 6, DC (C$_{48}$H$_{20}$) has 20 outer (solo and duo side-edged structures), 14 inner, and 14 center carbon sites, and we identify each carbon skeleton structure's effect on the hydrogenation processes. The reaction sites in symmetrical positions have similar reaction activity. According to that, 14 reaction pathways are considered and obtained. The reaction sites are highlighted with different colors and labeled: solo (purple, 1), duo (red, 2, 3, 4, and 5), inner (a, b, c, and d), and center (A, B, C, D, and E).

The binding energies of the mono-hydrogenated DC cations are given in the right panel of Figure 6. These 14 mono-hydrogenated DC cations are divided into three hydrogenation groups: outer edge, inner edge, and center edge. The first group is for the outer carbon edge, and we can see that all the reaction pathways are exothermic. For solo side-edged carbon sites, with -2.08 eV for [C$_{48}$H$_{20}$ + H(1)]$^+$; For duo side-edged carbon sites, with -1.74, -1.47, -1.46, and -1.77 eV for [C$_{48}$H$_{20}$ + H(2/3/4/5)]$^+$; The second group is for the inner carbon edge, we can see the reaction pathway is exothermic. With -1.04, -0.80, -1.16, and -0.86 eV for [C$_{48}$H$_{20}$ + H(a/b/c/d)]$^+$. The third group is for the center carbon edge, and we can see that all the reaction pathways are exothermic. With -1.06, -0.70, -1.02, -0.67, and -0.75 eV for [C$_{48}$H$_{20}$ + H(A/B/C/D/E)]$^+$, respectively.

The exothermic energy for the formation of [C$_{48}$H$_{21}$]$^+$, in the range of 0.67-2.08 eV, is relatively higher and can stabilize the whole molecules. The ion-molecule reactions between ovalene cations and H-atoms readily occur, and so we propose that [C$_{48}$H$_{21}$]$^+$ formed in the lab are a mixture with different possible isomers.

\begin{table*}
      \centering
      \caption{The theoretical calculation results for the bay region and non-bay region structures. Using ovalene ($\rm{C_{32}H_{14}}$) and periflanthene ($\rm{C_{32}H_{16}}$) as a comparable example. The doublet spin multiplicity is considered in all calculations in this table.}
      \label{tab:chap:table4}
      \renewcommand\arraystretch{3.0}%
      \setlength{\tabcolsep}{1mm}
      \begin{tabular}[c]{|c|c|c|c|c|}
             \hline
             ~~edge structure and sites diagram~~ & as the first H added & E(eV) & as the third H added & E(eV)\\ \hline
             \multirow{2}{*}{\begin{minipage}[b]{0.25\columnwidth}
                           \centering
                           \raisebox{-.5\height}{\includegraphics[width=\linewidth]{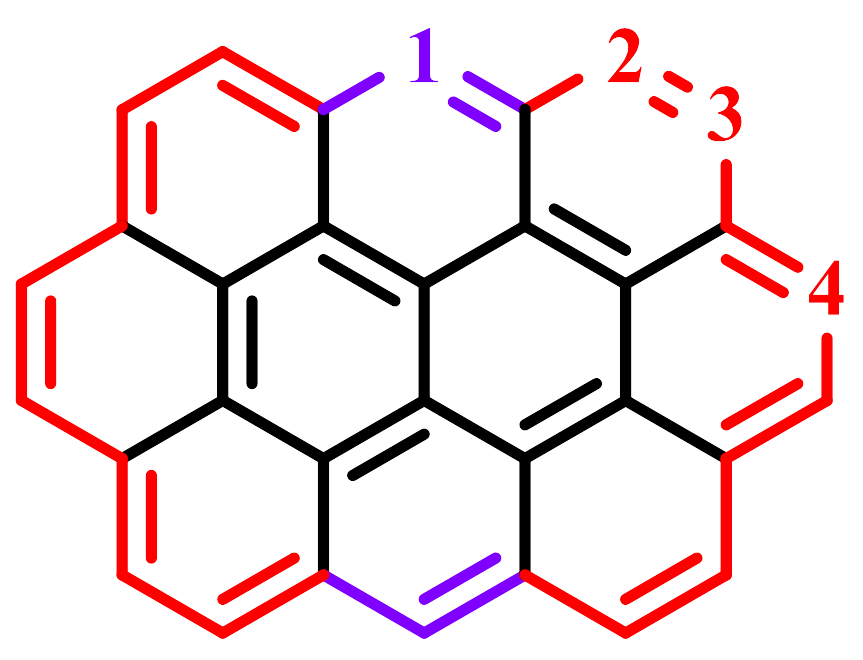}}
                            \raisebox{-.5\height}{Ovalene, non-bay region}
             \end{minipage}}
         & $\rm{[C_{32}H_{14}+H(1)]}^+$ & -2.31 & $\rm{[C_{32}H_{14}+2H(2,3)+H(1)]}^+$ & -2.00 \\ \cline{2-5}
             ~ & $\rm{[C_{32}H_{14}+H(4)]}^+$ & -1.81 & $\rm{[C_{32}H_{14}+2H(2,3)+H(4)]}^+$ & -1.63 \\ \cline{1-5}

             \multirow{2}{*}{\begin{minipage}[b]{0.38\columnwidth}
                           \centering
                           \raisebox{-.5\height}{\includegraphics[width=\linewidth]{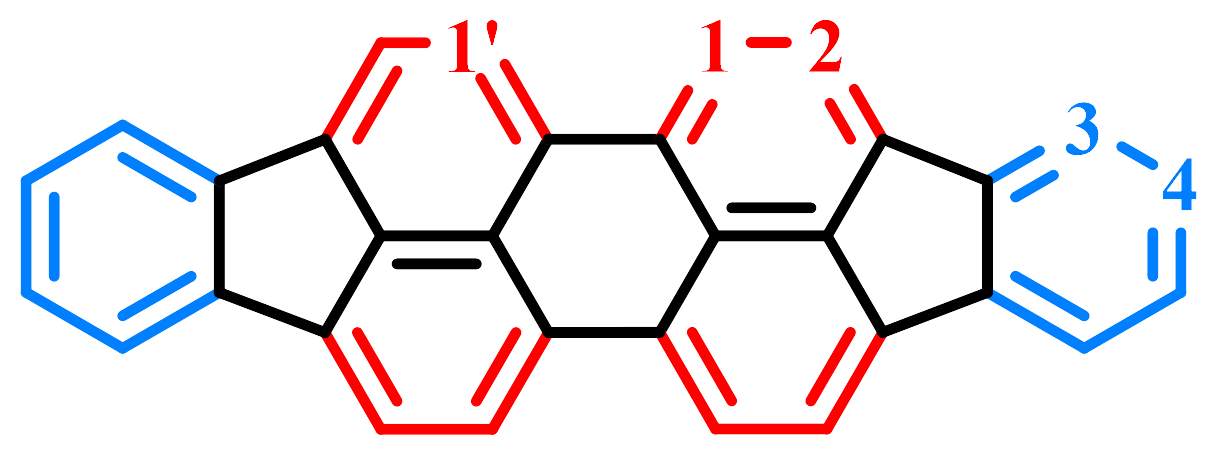}}
                                \raisebox{-.5\height}{periflanthene, with bay region}
             \end{minipage}}
         & $\rm{[C_{32}H_{16}+H(1')]}^+$ & -1.78 & $\rm{[C_{32}H_{16}+2H(1,2)+H(1')]}^+$ & -1.96 \\ \cline{2-5}
             ~ & $\rm{[C_{32}H_{16}+H(3)]}^+$ & -2.02 & $\rm{[C_{32}H_{16}+2H(1,2)+H(3)]}^+$ & -1.92 \\ \cline{1-5}
             \hline
      \end{tabular}
\end{table*}

\subsection{Calculation for the subsequent hydrogenation pathways}

For the subsequent hydrogenation pathways of the hydrogenated PAH cations, based on the molecular structural symmetry and the previous theoretical calculation results, we assume that further H-atom attachment and abstraction tend to occur on the outer carbon sites. In addition, the subsequent hydrogenation of hydrogenated PAH cations has many possible reaction pathways; for simplicity, we only chose the subsequent react carbon sites that are close to the already hydrogenated carbon sites \citep{pet95,lep01,caz16,zhang23}.

\subsubsection{The hydrogenation pathways for the hydrogenated TNPP cations ([C$_{40}$H$_{19}$]$^+$) $+$ H}

For the subsequent hydrogenation pathways of the hydrogenated PAH cations, we take TNPP as a typical example to calculate the hydrogenation processes. As shown in Table 1, TNPP has all kinds of CH groups, and the different edge structures are highlighted with different colors, which allows us to identify the edge effect on the hydrogenation processes for such a molecular geometry. Based on the obtained calculation results of hydrogenated TNPP cations ([C$_{40}$H$_{19}$]$^+$) $+$ H-atoms in Figure 1. For simplicity, to the subsequent adduct reaction pathways, hydrogenated TNPP cations ([C$_{40}$H$_{19}$]$^+$) $+$ H-atoms are calculated, and the calculation results are presented in Table 2.

Two types of reaction pathways are obtained: one reacts with its neighbor carbon sites (CH group) that add the H-atoms directly and form a CH$_2$ unit; the other one reacts with the hydrogenated group (CH$_2$) the CH unit is newly formed, and H$_2$ is formed as a secondary product.

To the solo edge, only one reaction pathways is obtained, [C$_{40}$H$_{18}$ + H(3)]$^+$ + H $\rightarrow$ [C$_{40}$H$_{18}$]$^+$ + H$_2$, the exothermic energy is 2.09 eV; To the duo, trio, and quarto edge, two type reaction pathway obtained, one is add the second H-atoms in its neighbor carbon sites, the other one is form H$_2$. We can see that all the reaction pathways are exothermic. The exothermic energy for the formation of [C$_{40}$H$_{20}$]$^+$, in the range of 1.60-2.70 eV, is relatively higher and can stabilize the whole molecules. The ion-molecule reactions between C$_{40}$H$_{19}$ cations and H-atoms readily occur, and so we propose that [C$_{40}$H$_{20}$]$^+$ formed in the lab are a mixture with different possible isomers.

H$_2$ is efficiently formed as a secondary product, and the exothermic energies for the dehydrogenated pathway are around $\sim$ 2.3 eV. Given the results presented in this work, the potential for large PAH cations to act as catalysts for molecular H$_2$ formation is confirmed \citep{bos12,thr19,pan19}. In addition, except for the trio edge, all the exothermic energies for the second are higher than the first one. The higher exothermic energies suggest that after adding the first H-atom, its neighbor carbon site becomes more chemically active. The trio edge structure, mainly due to the hydrogenation of the middle trio carbon site, destroys the aromaticity of the whole C-ring.

\subsubsection{The hydrogenation pathways for the hydrogenated PAH cation with the bay region and non-bay region structures}

For PAHs with the bay region and non-bay region structures, we take ovalene ($\rm{C_{32}H_{14}}$), periflanthene ($\rm{C_{32}H_{16}}$) as a comparable example with the same size to calculate the hydrogenation processes. As shown in Table 1, ovalene has a non-bay region structure, and periflanthene has six bay-region structures. This allows us to identify the bay-region effect on the hydrogenation processes for such a molecular geometry.

For simplicity, the calculation results are presented in Table 3. To ovalene cations, two reaction pathways are calculated, [C$_{32}$H$_{14}$ + 2H(2,3)]$^+$ + H $\rightarrow$ [C$_{32}$H$_{14}$ +3H(2,3,1)]$^+$ and [C$_{32}$H$_{14}$ + 2H(2,3)]$^+$ + H $\rightarrow$ [C$_{32}$H$_{14}$ + 3H(2,3,4)]$^+$. We can see the reaction pathway is exothermic, with -2.00 and -1.63 eV. To periflanthene cations, two reaction pathways are calculated, [C$_{32}$H$_{16}$ + 2H(1,2)]$^+$ + H $\rightarrow$ [C$_{32}$H$_{16}$ +3H(1,2,1')]$^+$ and [C$_{32}$H$_{16}$ + 2H(1,2)]$^+$ + H $\rightarrow$ [C$_{32}$H$_{16}$ + 3H(1,2,3)]$^+$. We can see the reaction pathway is exothermic, with -1.96 and -1.92 eV.

We can see, when compared with the first H-atoms, the exothermic energies of hydrogenation reaction pathways on the bay or non-bay region of hydrogenated cations located in a similar range and have no differences.

\begin{figure*}
      \subfigure{
             \begin{minipage}[t]{1\linewidth}
                    \centering
                    \includegraphics[width=5.2in]{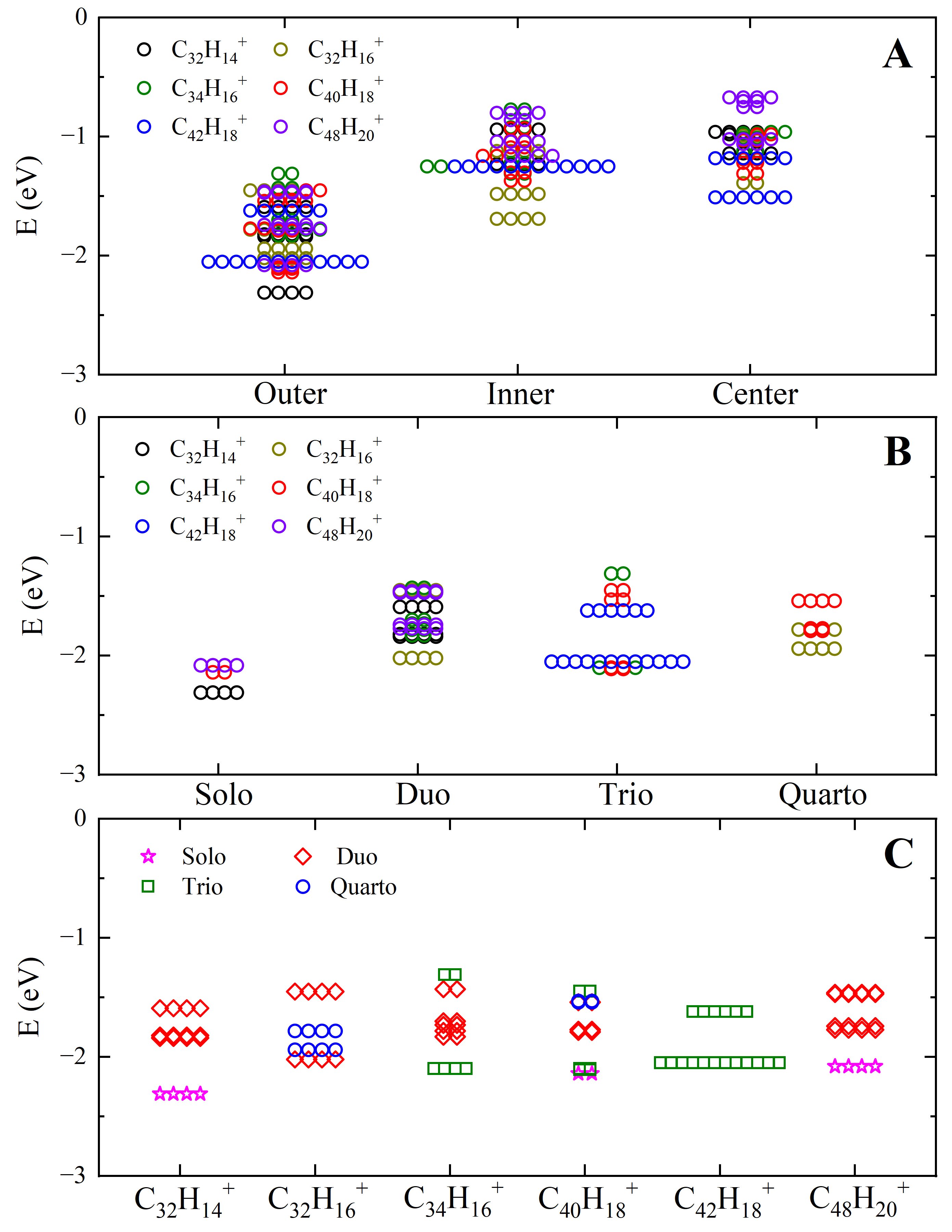}
             \end{minipage}%
      }%
      \centering
      \caption{The influencing factors that affect the hydrogenation chemical reactivity: panel (A): carbon skeleton structure's effect: outer, inner, and center; panel (B): side-edged structures' effect: solo, duo, trio, and quarto; panel (C): molecular size's effect.
      }
      \vspace{0.02cm}
      \label{fig7}
\end{figure*}

\begin{figure*}
      \centering
      \includegraphics[width=0.9\linewidth]{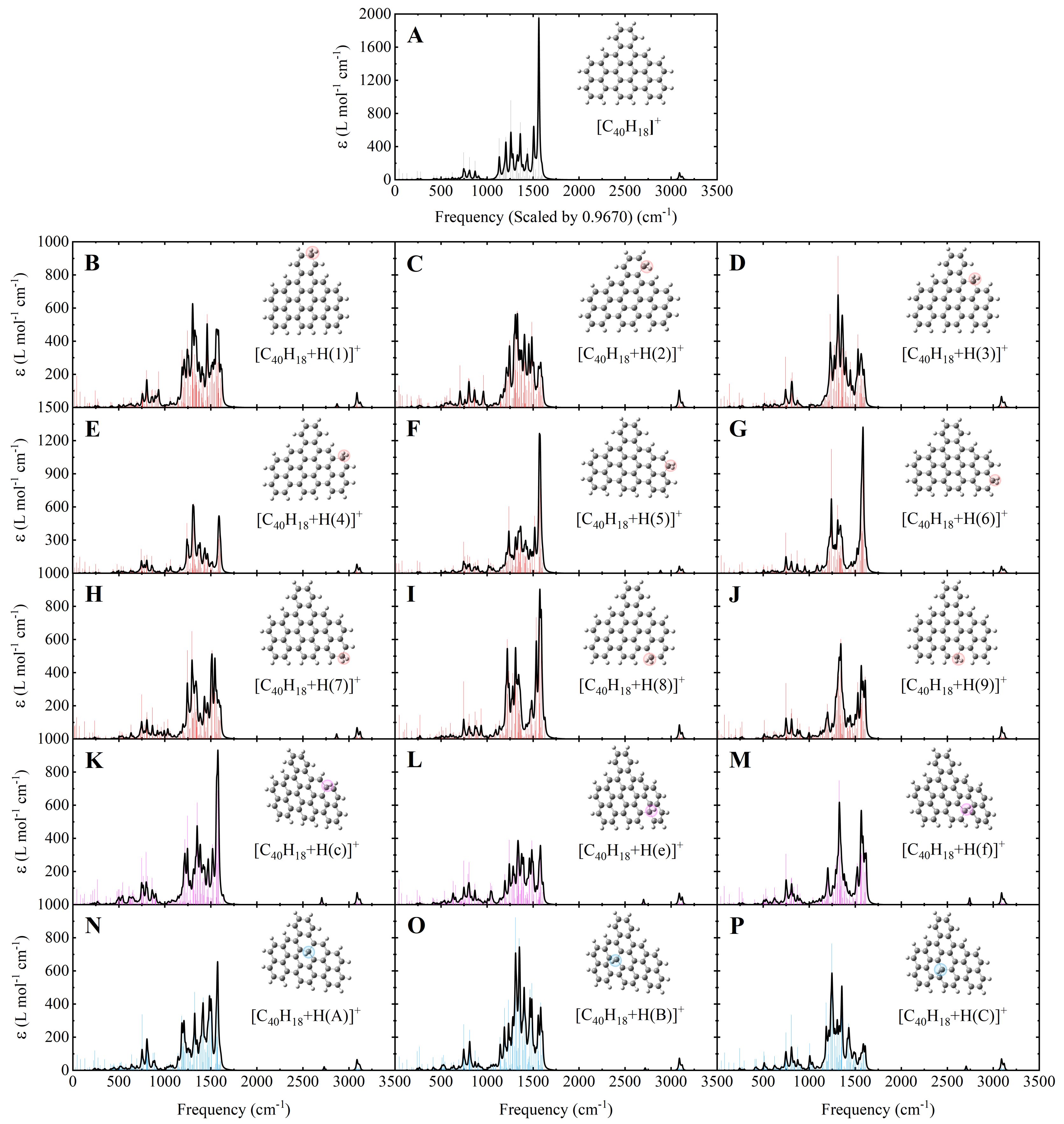}
      \caption{Computed vibrational normal modes for the mono-hydrogenated TNPP cations: panel (A) is the spectrum of [C$_{40}$H$_{18}$]$^+$; panel (B-J) is the spectrum of [C$_{40}$H$_{18}$+H]$^+$ (H atom added to outer carbon sites); panel (K-M) is the spectrum of [C$_{40}$H$_{18}$+H]$^+$ (H atom added to inner carbon sites); panel (N-P) is the spectrum of [C$_{40}$H$_{18}$+H]$^+$ (H atom added to center carbon sites). The vibrational band positions are scaled by a constant factor of 0.9670.}
      \vspace{0.02cm}
      \label{fig8}
\end{figure*}

\begin{figure*}
	\centering
	\includegraphics[width=0.9\textwidth]{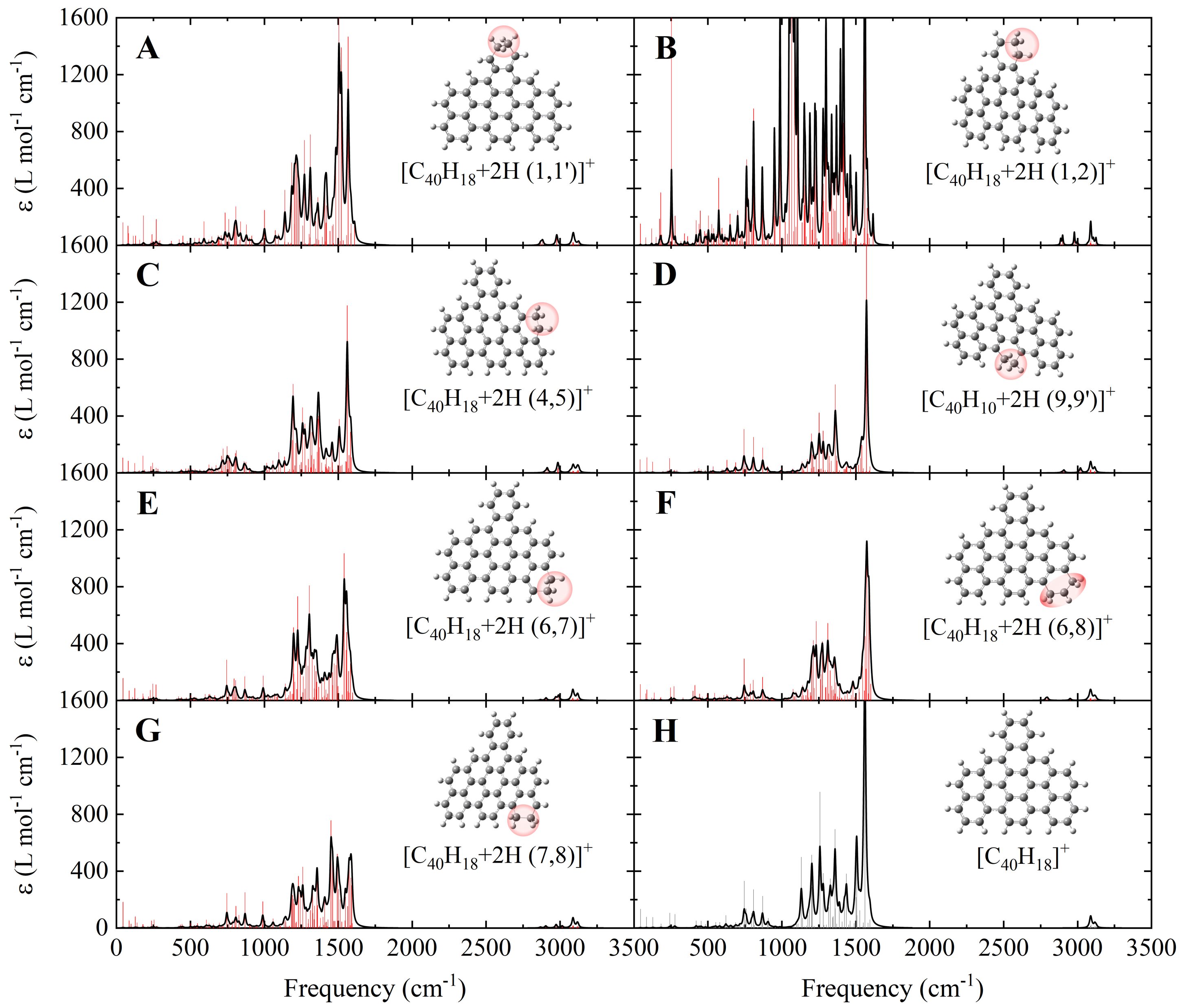}
	\caption{Computed vibrational normal modes for the di-hydrogenated TNPP cations ([C$_{40}$H$_{18}$+2H]$^+$). The vibrational band positions are scaled by a constant factor of 0.9670.
	}
	\label{fig9}
\end{figure*}

\subsection{Results discussion}
\label{sec:theoretical}

Overall, the results obtained from the theoretical calculation are consistent with the experimental results. The obtained exothermic energy for each hydrogenation reaction pathway is relatively high, which is consistent with the results we observed in our experiments that the ion-molecule collision reactions between PAH cations and H-atoms readily occur. From the calculation results on the hydrogenation processes of the hydrogenated PAH cations, we conclude that adding H-atoms to the carbon skeleton of PAH cations is a relatively independent event. As these reaction pathways are relatively independent and random events, the chance of their occurrence is entirely determined by the already hydrogenated carbon sites and the remaining hydrogenated carbon sites on the carbon skeleton of large cationic PAHs \citep{caz16,caz19,zhang23}. In addition, the competition between hydrogenation and dehydrogenation is confirmed that H$_2$ is efficiently formed as a secondary product. Given the results presented in this work, the potential for large PAH cations to act as catalysts for molecular H$_2$ formation is confirmed \citep{bos12,thr19,pan19}.

From the theoretical calculation, bonding ability plays an important role in the hydrogenation processes. Accordingly, some hydrogenation reaction pathways have higher exothermic energies, which means that these hydrogenation processes are more likely to occur, i.e., the carbon sites for hydrogenation have relatively high chemical reactivity. In the following subsections, we will discuss the factors that affect the chemical reactivity of these carbon reaction sites, including the effect of carbon skeleton structure, the side-edged structure, the molecular size, the bay region structure, and the neighboring hydrogenation, as presented in Figure 7.

\subsubsection{Carbon skeleton structure's effect: outer, inner, and center}

The exothermic energies of all these carbon sites for the hydrogenation are presented in Figure 7(A) as a function of carbon skeleton structures. We can see the exothermic energies of hydrogenation reaction pathways on the different types of carbon skeleton (e.g., outer, inner, and center carbon sites) have some differences: the exothermic energies of inner and center carbon sites are similar that located in the range of 0.6-1.6 eV, while the exothermic energies of outer carbon sites are mainly located in the range of 1.3-2.3 eV. Although there is some overlap in their energy ranges, the higher exothermic energies of outer than inner and center suggest that outer carbon sites have relatively higher chemical reactivity than the inner and center carbon sites, i.e., the outer carbon sites will more easily be hydrogenated. The H atom attachment and abstraction tend to occur on the outer carbon sites according to previous investigations \citep{eli09,caz16}.

However, in terms of the total number of carbon sites, the sum of inner and center carbon sites is more than that of outer, especially as the PAH becomes larger and more compact, the proportion of inner and center carbon sites will increase, i.e., the proportion of outer carbon sites in the total number will affect the collision reaction probability.

\subsubsection{Side-edged structure's effect: solo, duo, trio, and quarto}

The exothermic energies of all these outer side-edged carbon sites for the hydrogenation are presented in Figure 7(B) as a function of side-edged structures. We can see the exothermic energies of hydrogenation reaction pathways on the different types of side-edged structures (e.g., solo, duo, trio, and quarto) have some differences: the exothermic energies of solo are located in the range of 2.1-2.3 eV; the exothermic energies of duo are mainly located in the range of 1.3-2.0 eV; the exothermic energies of trio are mainly located in the range of 1.2-2.1 eV; the exothermic energies of quarto are mainly located in the range of 1.7-1.9 eV.

Although there is some overlap in their energy ranges, the higher exothermic energies of solo carbon sites are higher than others, and the solo carbon sites have relatively higher chemical reactivity than others. The exothermic energies of duo and quarto carbon sites are quite similar, suggesting that duo and quarto carbon sites have relatively similar chemical reactivity. The exothermic energies of trio carbon sites are scattered in a relatively large range that the middle carbon sites are $\sim$ 0.5 eV smaller than the side, which suggests that side carbon sites have relatively higher chemical reactivity than the middle position of trios. Based on that, the solo carbon sites and the side carbon sites of the trio edge will more easily be hydrogenated. However, in terms of each PAH, the proportion of each side-edged structure in the total number will affect the collision reaction probability.

\subsubsection{Molecular size's effect}

The exothermic energies of all these outer side-edged carbon sites for the hydrogenation are presented in Figure 7(C) as a function of molecular size. The exothermic energies of hydrogenation reaction pathways on different sizes (e.g., from 32 to 48 C-number) are located in a similar range and have no differences. PAHs with different molecular sizes have a relatively similar chemical reactivity, i.e., no molecular size effect on the hydrogenation processes. The no molecular size effect conclusion is working in the size range (32-48 C-atoms) that is presented here, but it may not work in other molecular size ranges.

\subsubsection{Five- and six-membered C-ring structure's effect}

As presented in Figure 2 and Figure 3, ovalene and periflanthene have 32 C-atoms, but their molecular structure is different. Ovalene is a pericondensed PAH with six-member C-ring, and periflanthene is a catacondensed PAH with two five-membered C-ring (only for the inner and center carbon sites). From the obtained calculation results, we can see the exothermic energies of hydrogenation reaction pathways on five-membered C-ring and six-membered C-ring have some differences: with -1.69 and -1.48 eV for [C$_{32}$H$_{16}$ + H(b/c)]$^+$, -1.39 eV for [C$_{32}$H$_{16}$ + H(B)]$^+$, respectively; while with -0.94 and -1.23 eV for [C$_{32}$H$_{14}$ + H(a/b)]$^+$,-0.98, -1.14 and -0.96 eV for [C$_{32}$H$_{14}$ + H(A/B/C)]$^+$, respectively. The five-membered C-ring carbon sites have relatively higher chemical reactivity than the six-membered C-ring.

\subsubsection{Neighboring hydrogenation's effect}

As presented in Table 2, from our calculations, except for the trio edge, all the exothermic energies for the second (in the neighbor carbon sites) are higher than the first one, i.e., the carbon site becomes more chemically active by its neighboring carbon site's hydrogenation. The trio edge structure, mainly due to the hydrogenation of the middle carbon site, destroys the aromaticity of the whole C-ring. We note that we performed only some of these related theoretical calculations. When the amount of hydrogen addition is higher, some intramolecular interactions and other reaction formation pathways may also exist.

\subsubsection{Bay region structure's effect}

From previous work \citep{cas18}, the bay and non-bay region structure have a very important impact on dehydrogenation. However, as presented in Table 3, from our calculations, the exothermic energies of hydrogenation reaction pathways with bay or non-bay structures have no differences. PAHs with bay or non-bay structures have a similar chemical reactivity, i.e., no bay region structure's effect on the hydrogenation processes. These reaction pathways are relatively independent and random events, and the chance of their occurrence is entirely determined by the already hydrogenated carbon sites and the remaining hydrogenated carbon sites on the carbon skeleton of large cationic PAHs \citep{caz16,zhang23}.

\section{IR spectra of hydrogenated PAH cations}
\label{sec:ir}

These possible newly formed hydrogenated PAH cations obtained above may also be the candidates of the observed IR interstellar bands \citep{sch93}. They may exist in the same interstellar area through an evolutionary ion-molecular reaction network between PAHs and H-atoms that may contribute to the observed interstellar spectrum, which motivates the spectroscopic studies \citep[and references therein]{pee04,tie13}. Based on the calculation results and to obtain the IR spectra of these molecules after the hydrogenation processes, the IR spectra of these molecules were obtained. For these newly formed molecules, the calculations were performed with the 6-311++G(d, p) basis set. The vibrational band positions are scaled by a uniform factor of 0.9670, and the black line is the spectra simulated by Gaussians with a full-width half-maximum of 8 cm$^{-1}$ \citep{boe2014}.

\subsection{IR spectra of mono-hydrogenated PAH cations}

The computed vibrational normal modes for the mono-hydrogenated TNPP cations are presented in Figure 8, in the range of 0$-$3500 cm$^{-1}$: panel (A) is the IR spectrum of [C$_{40}$H$_{18}$]$^+$ (without hydrogenation); panel (B-J, red) is the IR spectrum of [C$_{40}$H$_{18}$+H]$^+$ (H atom added to outer carbon sites); panel (K-M, pink) is the IR spectrum of [C$_{40}$H$_{18}$+H]$^+$ (H atom added to inner carbon sites); panel (N-P, blue) is the IR spectrum of [C$_{40}$H$_{18}$+H]$^+$ (H atom added to center carbon sites). For further comparison, the IR spectra of other mono-hydrogenated PAH cations are provided in Appendix B (Figures B1-B5).

The IR spectra of [C$_{40}$H$_{18}$+H]$^+$ are very complex, and we only conclude two characteristic peaks: 3.4-3.5 $\mu$m is the stretch vibration of CH$_2$ when the H-atoms are added to the outer carbon sites of TNPP cations; 3.6-3.7 $\mu$m is the stretch vibration of CH when the H-atoms are added to the inner and center carbon sites of TNPP cations. As can be seen in Figure 8, the spectral intensity of hydrogenated TNPP cations becomes weaker when compared to that of C$_{40}$H$_{18}$$^+$ (without hydrogenation) (e.g., the strength of C-C stretch).

\subsection{IR spectra of di-hydrogenated PAH cations}

The computed vibrational normal modes for the di-hydrogenated TNPP cations are presented in Figure 9, in the range of 0$-$3500 cm$^{-1}$. As can be seen in Figure 9, 3.3-3.6 $\mu$m are the characteristic stretch vibrations of CH$_2$ when the TNPP cations attached with 2 extra H atoms, which is a wider range than that of mono-hydrogenated TNPP cations. Moreover, the spectral intensity of di-hydrogenated TNPP cations (except [C$_{40}$H$_{18}$+2H(1,2)]$^+$) becomes weaker when compared to that of C$_{40}$H$_{18}$$^+$ (without hydrogenation). It has no significant differences when compared to that of [C$_{40}$H$_{18}$+H]$^+$ (mono-hydrogenated TNPP cations) (e.g., the strength of C-C stretch).

In general, the IR spectra of the TNPP cations are relatively similar to the newly formed hydrogenated TNPP cations. Some new vibrational modes were preserved, while others were not, and some newly vibrational modes were formed, and the vibrational peaks were variable due to the influence of their surrounding environment. These obtained spectra reveal the complexity of molecule structure and the complexity of their infrared spectra, which also illustrates the important role of the complex organic molecules in the carbon skeleton and the connection type of clusters.

\section{Astronomical implications}
\label{sec:astro}

As a complex molecular system, PAH compounds have very complex molecular evolution characteristics and spatial distribution in interstellar space \citep[and references therein]{tie13}. In the chemical evolution network, PAH molecules are affected and constrained by the interstellar UV radiation, the flux of H-atoms, hydrogen ions, and other coexisting molecules during the different interstellar evolution periods \citep{tie13,zhang23}. The evolution of PAH molecules in response to H atom bombardment is very complex. In the present paper, we demonstrate the processes taking place in PAH molecules in response to H atom bombardment from different perspectives, including in experiments and quantitative calculations. In contrast to our laboratory studies, hydrogenation of PAHs in PDRs is similar to that, mainly due to H-atom bombardment in the gas phase. Hence, our experiments can be directly applied to understand the hydrogenation processes of PAHs in space.

Six large, astronomically relevant PAHs are selected in this work. These selected six PAHs are used here as prototypical examples for large(r) PAHs \citep{and2015,cro16}, and their size, from 32 to 48 C-number, that lies within the astrophysically relevant range \citep{kok08,ste11,gre11}. The experimental results show that the collision reactions between PAH cations and H-atoms can efficiently occur in the gas phase, and the hydrogenated PAH cations are formed through a sequential-step pathway. Unlike the photo-dehydrogenation processes that lead to the predominance of even-mass fragments \citep{zhen2014}, there are no even-odd hydrogenated mass patterns observed in the gas-phase hydrogenation processes.

The results of the theoretical calculations are consistent with the experimental results. From the theoretical calculation, the exothermic energy for each reaction pathway is relatively higher, meaning that the ion-molecule reactions between the PAH and H atoms are mainly one-off collision reactions. The molecular structure of hydrogenated PAH cations is diverse. The relative chemical activity is very important in the ion-molecular reaction formation process and reaction evolution network, which affects the abundance ratio of formed species \citep{jag09,mon13,zhen2019}. From the obtained results, we infer that the formation of aliphatic C-H bonds plays an important role in the chemical evolution process of PAHs.

Furthermore, the calculated exothermic energy for each formation steps are very high. The exothermic energy can quickly be transferred as the vibrational energy of ground states to "heat" the molecules, resulting in an increase in the temperature of molecules \citep{tie13}. In our experimental condition, the exothermic energy is mainly released through collision with He atoms (buffer gas); IR fluorescence may not play an important role in the exothermic energy release processes. In contrast to our laboratory studies, in the astrophysical environments, due to the low density of molecules, the exothermic energy is mainly released through IR fluorescence. However, irrespective of which energy release pathways are involved, it will always leave the species with lower vibrationally excited from which it relaxes either through collision relaxation or IR fluorescence.

The AIBs, which are key PDR signatures in the near- and mid-IR regions, are mainly attributed to PAHs and related species that are heated by the absorption of UV photons \citep{leg84,all85}. Therefore, most of the galactic and extragalactic observations obtained with the newly launched infrared James Webb Space Telescope (JWST) will encompass PDR emissions and will help astrochemical researchers to elucidate the distribution and molecular and spatial evolution characteristics of PAH molecules in interstellar environments \citep{ber22}. Especially for these IR spectra of the hydrogenated PAH cations obtained in this work, their spectral profile contributions and the spectral features may offer some insights for the tentative detection in the ISM.

\section{Conclusions}
\label{sec:conc}

Overall, combining experiments with quantum chemical calculations, the hydrogenation processes between six large, astronomically relevant PAH cations and H-atoms are studied in this work. The hydrogenated PAH cations are efficiently formed through an ion-molecule collision reaction in the gas phase. Through theoretical quantum calculations, the structure of newly formed hydrogenated PAH cations and the bonding energies of their reaction pathways, together with the IR spectra, are obtained.

The effect of carbon skeleton structure, the side-edged structure, the molecular size, the five- and six-membered C-ring structure, the bay region structure, and the neighboring hydrogenation are discussed based on the calculation results:
\begin{enumerate}
       \item The outer carbon sites have relatively higher chemical reactivity than the inner and center carbon sites, i.e., the outer carbon sites will be hydrogenated more easily.
       \item The solo and the side carbon sites of the trio edges will more easily be hydrogenated.
       \item PAHs with different molecular sizes have a relatively similar chemical reactivity, i.e., no molecular size effect on the hydrogenation processes.
       \item The five-membered C-ring carbon sites have relatively higher chemical reactivity than the six-membered C-ring.
       \item The carbon site becomes more chemically active by its neighboring carbon site's hydrogenation.
       \item PAHs with bay or non-bay structures have a similar chemical reactivity, i.e., large PAHs do not have the bay region structure's effect on the hydrogenation processes.
\end{enumerate}

The results we obtained once again validate the complexity of interstellar molecules, and the evolution of PAH molecules under the constraints of the factors of H-atoms is very complex. Hence, if these PAHs are present in space, the formation of hydrogenated PAH cations could produce an extended family of large molecules. As part of the coevolution of interstellar chemistry, the hydrogenation and dehydrogenation reaction modes should be considered where these PAHs are located when further reacting with H-atoms in the interstellar environment, which provides data support and direction for our future treatment of interstellar evolution simulations, especially for the interstellar coevolution simulations.

Furthermore, these hydrogenated PAH cations may also provide candidates of interest for the IR interstellar bands. From the theoretical results, we can see that more aliphatic C-H bonds formed with its new vibrational modes. Therefore, their role in the spectral profile contributions and the spectral features for tentative detection in the ISM is warranted \citep{tie08,ber22,cho23}.

\section*{Acknowledgements}

This work is supported by the National Natural Science Foundation of China (NSFC, Grant No. 12333005, 12122302, and 12073027). Theoretical calculations were performed at the Supercomputing Center of the University of Science and Technology of China.

\section*{Data availability}

All data generated or analyzed during this study are included in this article.

\clearpage

\appendix

\section{Experimental methods}

Here, we provide a brief description of the experimental methods used in this study \citep{zhang23}. Neutral PAHs were converted into the gas phase by heating PAHs in powdered form (provided by Kentax, with a purity greater than 99.5 \%) in an oven. The temperature of oven is $\sim$ 473, 473, 503, 553, 573, and 603 K for ovalene ($\rm{C_{32}H_{14}}$), periflanthene ($\rm{C_{32}H_{16}}$), tri-benzo-peropyrene (TBP, $\rm{C_{34}H_{16}}$), tribenzo-naphtho-pero-pyrene (TNPP, $\rm{C_{40}H_{18}}$), hexa-benzo-coronene (HBC, $\rm{C_{42}H_{18}}$), and dicoronylene (DC, $\rm{C_{48}H_{20}}$), respectively. Subsequently, evaporated PAH molecules were ionized and transported into the ion trap via an ion-gate. During this procedure, helium gas was introduced continuously into the trap via a leaking valve to thermalize the ion cloud through collisions ($\sim$ 300 K). A high-precision delay generator (SRS DG535) was used to control the full-timing sequence.

A hydrogen atom beam source (HABS; MBE-Komponenten GmbH; \cite{tsc98}) was installed to produce the H atoms. The H-atoms were formed by cracking H$_2$ gas (with a purity greater than 99.99 \%) using a tungsten capillary at $\sim$ 2173 K. The chamber pressure during the H atom beam exposition was $\sim$ 1.2 * 10$^{-6}$ mbar with H$_2$ flowing through the HABS (the typical background pressure in the chamber was $\sim$ 6.0 * 10$^{-7}$ with helium gas). The working distance from the end of the tungsten capillary to the center of the ion trap was $\sim$ 5.0 cm. The H-atoms flux was incident into the ion trap through a 2.4 mm aperture in the ion trap ring electrode. Based on the operating conditions, the H-atoms were expected to have a flux of $\sim$ 4.0 * 10$^{14}$ H-atoms cm$^{-2}$s$^{-1}$, and the volume density of H-atoms is $\sim$ 5*10$^9$ cm$^{-3}$.

A delay generator controls the full-timing sequence. Our setup is operated with a frequency of 0.2, 0.1, 0.05, or 0.025 Hz, i.e., one complete measuring cycle lasts 5.0, 10.0, 20.0, or 40.0 s, respectively. In each experiment, the ion gate is opened, allowing the ion trap to fill with a certain amount of ions. Hydrogenated PAH cations are formed in the ion trap, and adduct formation presumably occurs. A negative square pulse is applied to the end cap at the end of each cycle, and the resulting mass fragments are measured.

\section{Vibration modes and corresponding infrared intensity of other hydrogenated PAH cations}

The IR spectra of other hydrogenated PAH cations are presented in Figures B1-B5, in the range of 0$-$3500 cm$^{-1}$. The vibrational band positions are scaled by a constant factor of 0.9670, and the black line represents the spectra simulated by Gaussians with a full width at half maximum of 8 cm$^{-1}$ \citep{boe2014}.

In Figure B1, the computed vibrational normal modes for the mono-hydrogenated ovalene cations ([C$_{32}$H$_{14}$+H]$^+$) are presented. In Figure B2, the computed vibrational normal modes for the mono-hydrogenated periflanthene cations ([C$_{32}$H$_{16}$+H]$^+$) are presented. In Figure B3, the computed vibrational normal modes for the mono-hydrogenated TBP cations ([C$_{34}$H$_{16}$+H]$^+$) are presented. In Figure B4, the computed vibrational normal modes for the mono-hydrogenated HBC cations ([C$_{42}$H$_{18}$+H]$^+$) are presented. In Figure B5, the computed vibrational normal modes for the mono-hydrogenated DC cations ([C$_{48}$H$_{20}$+H]$^+$) are presented.

\begin{figure*}
      \centering
      \includegraphics[width=6.0in]{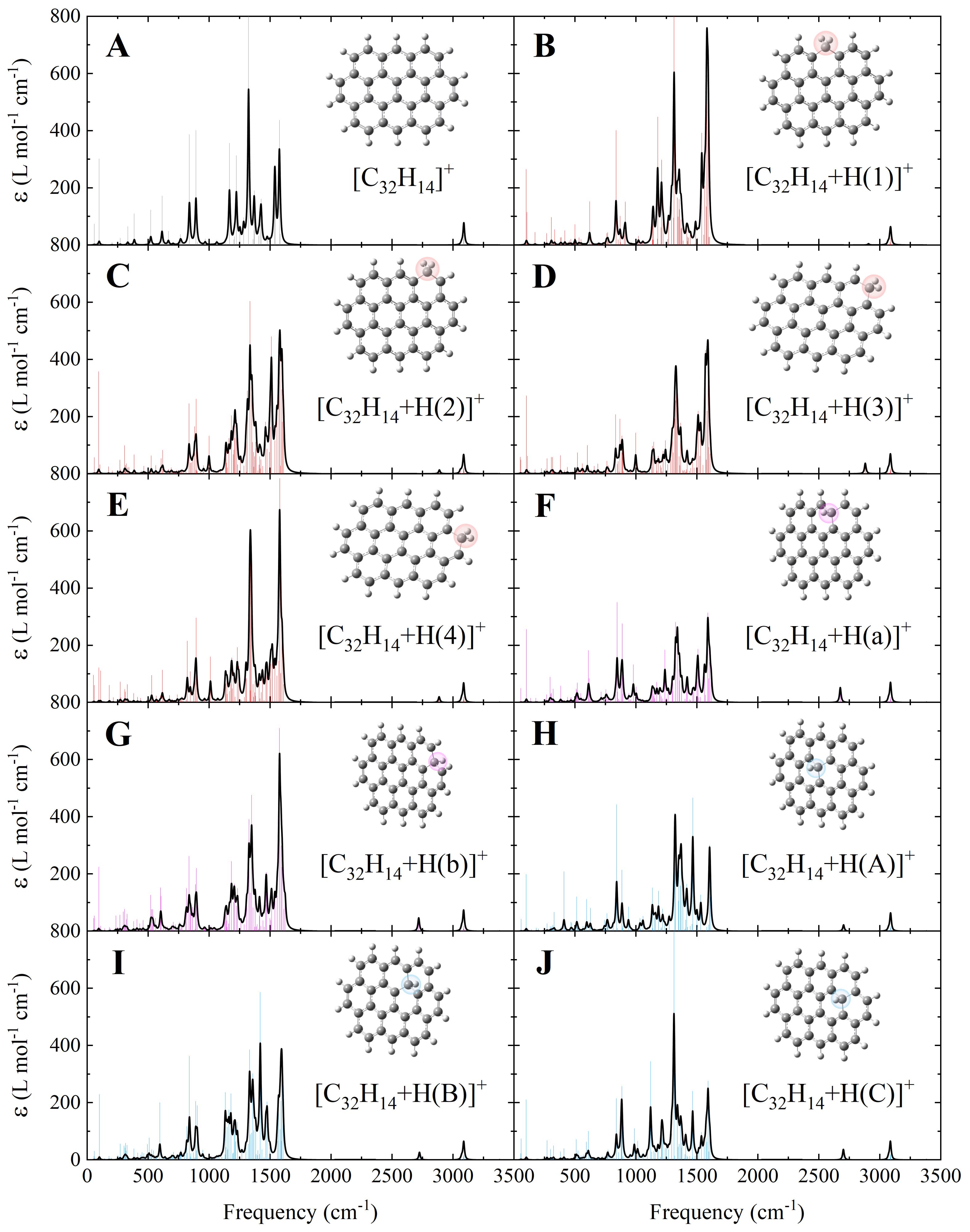}
      \caption{Computed vibrational normal modes for the mono-hydrogenated ovalene cations: panel (A) is the spectrum of [C$_{32}$H$_{14}$]$^+$; panel (B-E) is the spectrum of [C$_{32}$H$_{14}$+H]$^+$ (H atom added to outer carbon sites); panel (F-G) is the spectrum of [C$_{32}$H$_{14}$+H]$^+$ (H atom added to inner carbon sites); panel (H-J) is the spectrum of [C$_{32}$H$_{14}$+H]$^+$ (H atom added to center carbon sites). The vibrational band positions are scaled by a constant factor of 0.9670.
      }
      \label{spec1}
\end{figure*}

\begin{figure*}
       \centering
       \includegraphics[width=6.0in]{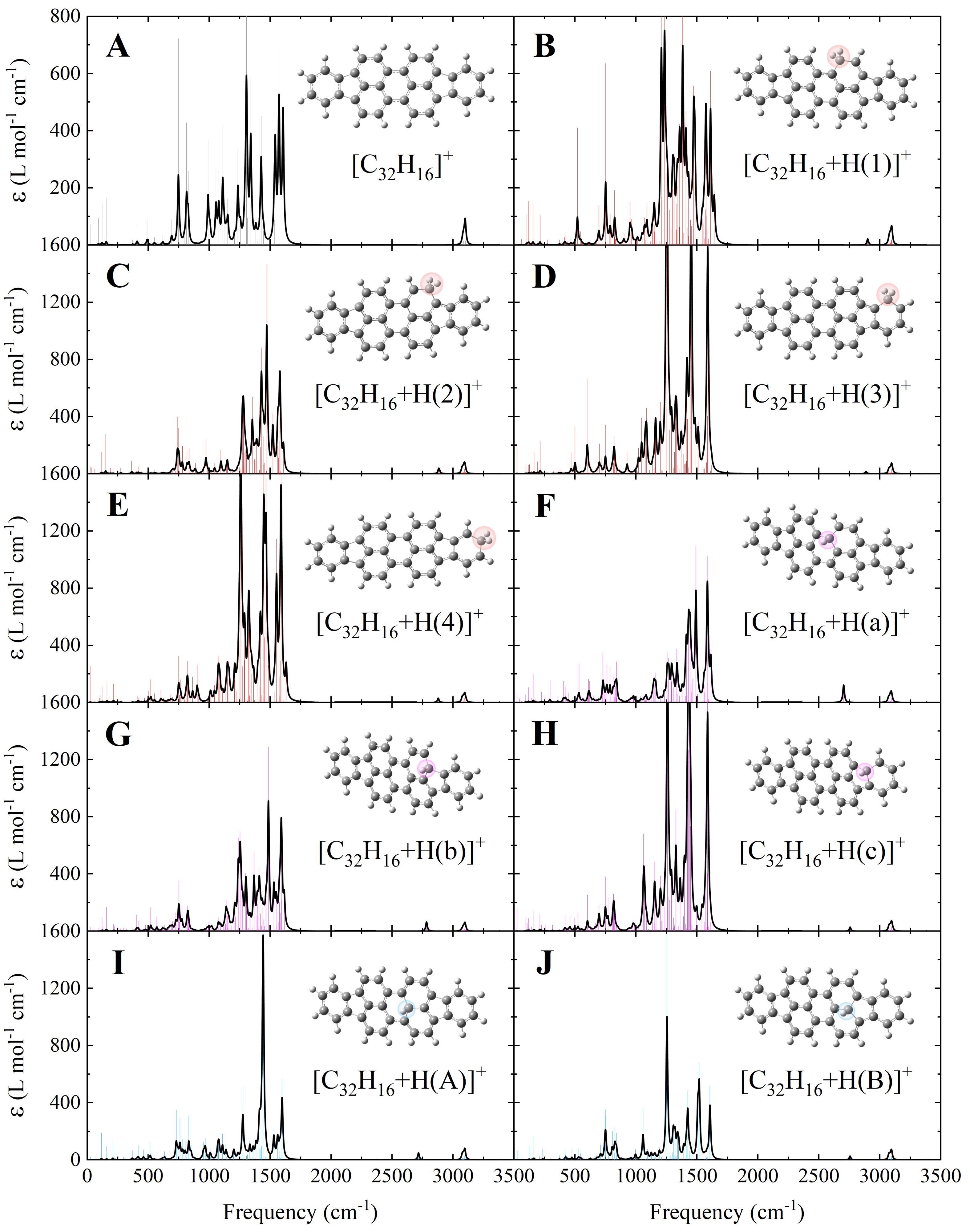}
       \caption{Computed vibrational normal modes for the mono-hydrogenated periflanthene cations: panel (A) is the spectrum of [C$_{32}$H$_{16}$]$^+$; panel (B-E) is the spectrum of [C$_{32}$H$_{16}$+H]$^+$ (H atom added to outer carbon sites); panel (F-H) is the spectrum of [C$_{32}$H$_{16}$+H]$^+$ (H atom added to inner carbon sites); panel (I-J) is the spectrum of [C$_{32}$H$_{16}$+H]$^+$ (H atom added to center carbon sites). The vibrational band positions are scaled by a constant factor of 0.9670.
       }
       \label{spec2}
 \end{figure*}

\begin{figure*}
      \centering
      \includegraphics[width=\textwidth]{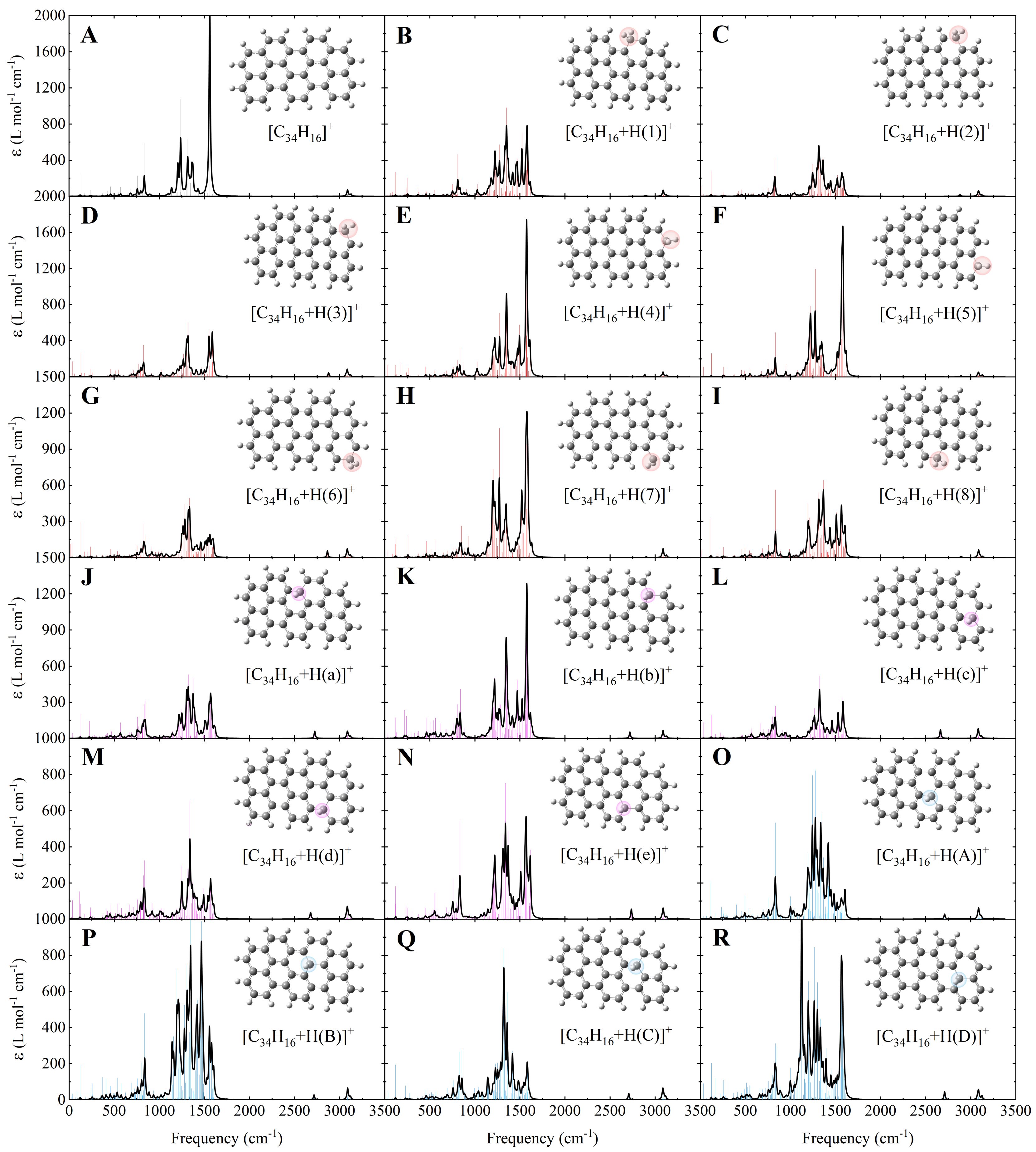}
      \caption{Computed vibrational normal modes for the mono-hydrogenated TBP cations: panel (A) is the spectrum of [C$_{34}$H$_{16}$]$^+$; panel (B-I) is the spectrum of [C$_{34}$H$_{16}$+H]$^+$ (H atom added to outer carbon sites); panel (J-N) is the spectrum of [C$_{34}$H$_{16}$+H]$^+$ (H atom added to inner carbon sites); panel (O-R) is the spectrum of [C$_{34}$H$_{16}$+H]$^+$ (H atom added to center carbon sites). The vibrational band positions are scaled by a constant factor of 0.9670.
      }
      \label{spec3}
\end{figure*}

\begin{figure*}
       \centering
       \includegraphics[width=\textwidth]{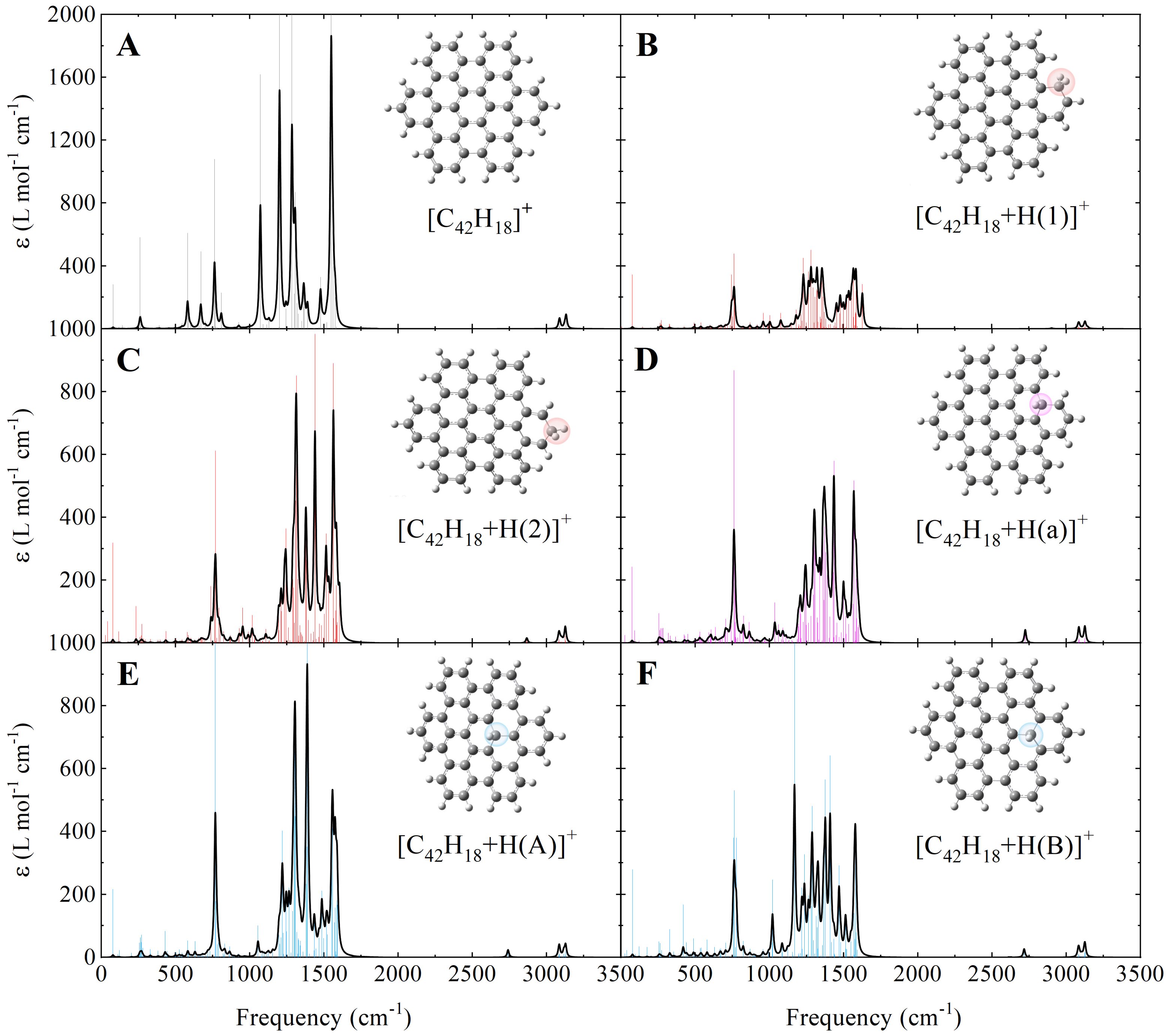}
       \caption{Computed vibrational normal modes for the mono-hydrogenated HBC cations: panel (A) is the spectrum of [C$_{42}$H$_{18}$]$^+$; panel (B-C) is the spectrum of [C$_{42}$H$_{18}$+H]$^+$ (H atom added to outer carbon sites); panel (D) is the spectrum of [C$_{42}$H$_{18}$+H]$^+$ (H atom added to inner carbon sites); panel (E-F) is the spectrum of [C$_{42}$H$_{18}$+H]$^+$ (H atom added to center carbon sites). The vibrational band positions are scaled by a constant factor of 0.9670.
       }
       \label{spec4}
 \end{figure*}

\begin{figure*}
      \centering
      \includegraphics[width=\textwidth]{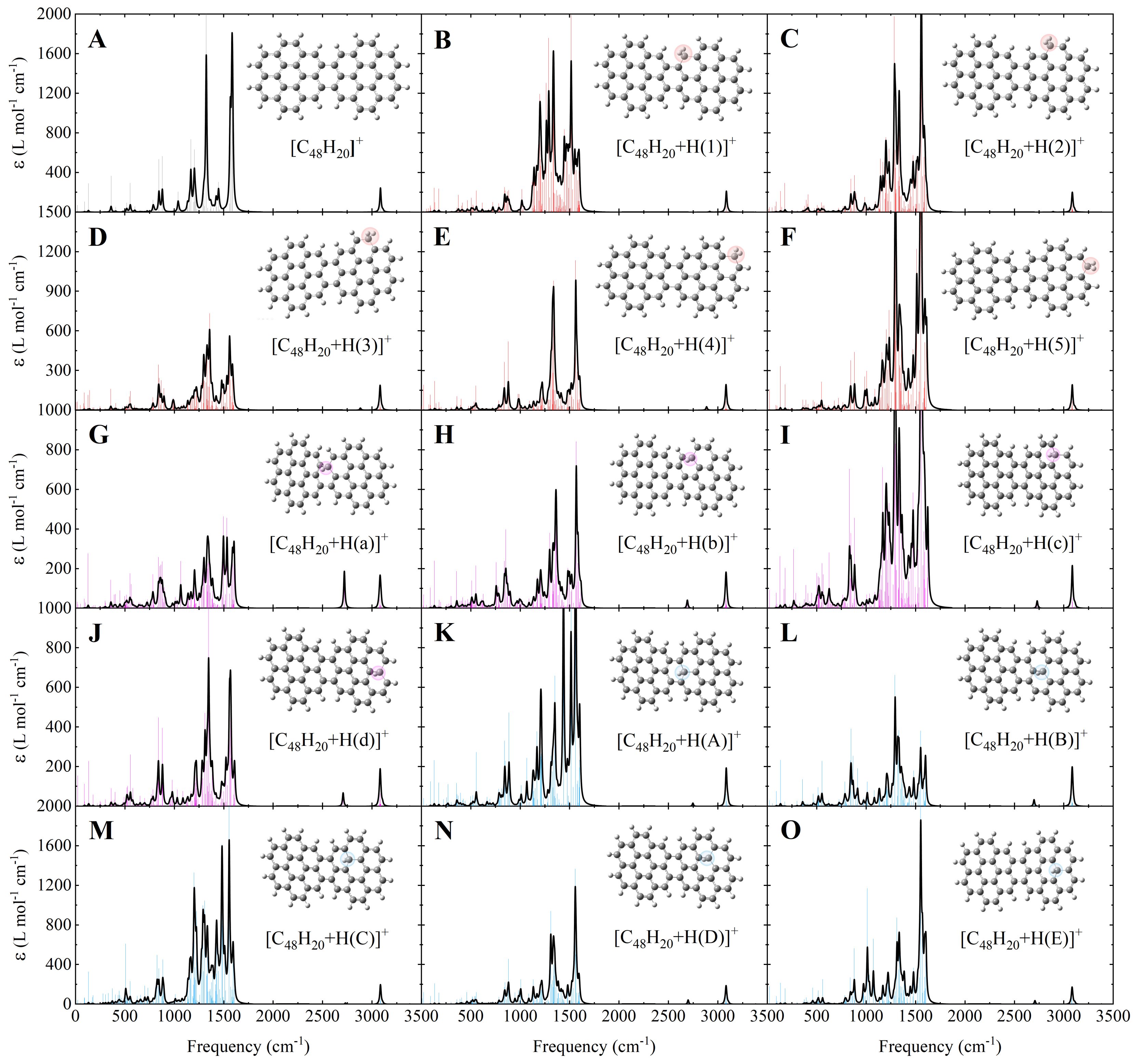}
      \caption{Computed vibrational normal modes for the mono-hydrogenated DC cations: panel (A) is the spectrum of [C$_{48}$H$_{20}$]$^+$; panel (B-F) is the spectrum of [C$_{48}$H$_{20}$+H]$^+$ (H atom added to outer carbon sites); panel (G-J) is the spectrum of [C$_{48}$H$_{20}$+H]$^+$ (H atom added to inner carbon sites); panel (K-O) is the spectrum of [C$_{48}$H$_{20}$+H]$^+$ (H atom added to center carbon sites). The vibrational band positions are scaled by a constant factor of 0.9670.
      }
      \label{spec5}
\end{figure*}

\end{document}